%% file: Llopis_et_al.tex
\documentclass[11pt]{amsart}
\usepackage[margin=1in]{geometry}% Just for this example\usepackage[margin=1in]{geometry}% Just for this example

\usepackage{cite,amssymb,graphicx,caption,lipsum}

\usepackage[utf8]{inputenc} % allow utf-8 input
\usepackage[T1]{fontenc}    % use 8-bit T1 fonts
\usepackage{hyperref}       % hyperlinks
\usepackage{url}            % simple URL typesetting
\usepackage{amsfonts}       % blackboard math symbols
\usepackage{lipsum}

\usepackage{abstract}

\usepackage{cite,amssymb,graphicx,caption,lipsum}
\usepackage{amsmath}
\usepackage{amsfonts}
\usepackage{enumerate}
\usepackage{epsfig}
\usepackage[dvipsnames]{xcolor}
\usepackage{psfrag}
\usepackage{hyperref}
\usepackage{babel}
\usepackage{lineno}
\usepackage{chemarr}
\usepackage{rotating}
\usepackage{mathtools}
\usepackage{graphicx}
\usepackage{float}
\usepackage{subfig}
\usepackage[font=small]{subcaption}
\usepackage{rotating}

\usepackage[colorinlistoftodos]{todonotes}
\usepackage{pgfplots}
\usepackage{stackengine}
\usepgfplotslibrary{fillbetween}
\pgfplotsset{compat=newest}
\usetikzlibrary{decorations.markings, patterns}

\usepackage[short,nodayofweek,level,12hr]{datetime}

\usepackage{tikz}
\usetikzlibrary{math}
\usepackage{pgfplots}
\usepackage{stackengine}
\usepgfplotslibrary{fillbetween}
\pgfplotsset{compat=newest}
\usetikzlibrary{decorations.markings, patterns}

\newtheorem{lemma}{Lemma}
\newtheorem{prop}{Proposition}
\newtheorem{remark}{Remark}
\newtheorem{defi}{Definition}

\definecolor{Bluish}{rgb}{0.,0.,0.5}
\definecolor{Reddish}{rgb}{0.5,0.,0.}

\def\eps{\varepsilon}

\def\QNC{$\mathrm{QNC}$}

\usepackage[colorinlistoftodos]{todonotes}
\usepackage[normalem]{ulem}

\definecolor{ColorTomasL}{rgb}{0.9,0.2,0.1}

\definecolor{ColorJS}{rgb}{0.9,0.,0.4}

\definecolor{ColorSanti}{rgb}{0.1,0.5,0.1}

\definecolor{ColorOriol}{rgb}{0.2,0.5,0.2}

\colorlet{Mag}{magenta!60}

\usepackage{graphicx}
\graphicspath{ {./figures/} }

\usepackage{enumerate}
\usepackage{color}
\usepackage{comment}
\usepackage{hyperref}

\def\ds{\displaystyle}

\newcommand{\R}{\ensuremath{\mathbb{R}}}

\renewcommand{\thefootnote}{\roman{footnote}}

\usepackage{lipsum}% http://ctan.org/pkg/lipsum
\makeatletter
\newcommand{\authorfootnotes}{\renewcommand\thefootnote{\@fnsymbol\c@footnote}}%
\renewcommand{\thefootnote}{\alph{footnote}}

\makeatother

\usepackage{authblk}
\begin{document}

\begin{center}
\LARGE
Global bifurcation in a virus, defective genomes, satellite RNAs tripartite system: breakdown of a coexistence quasi-neutral curve
 \par \bigskip

  \normalsize
  \authorfootnotes
  Oriol Llopis-Almela\footnote{O.~Llopis-Almela: ollopis@crm.cat (Corresponding author)}
  \textsuperscript{1},
  J.Tom\'as~L\'azaro\footnote{J.T.~L\'azaro: jose.tomas.lazaro@upc.edu}\textsuperscript{1,2,3,4}, 
  Santiago F. Elena\footnote{S.F.~Elena: santiago.elena@csic.es} \textsuperscript{5,6} and
  Josep Sardany\'es\footnote{ J. Sardany\'es: jsardanyes@crm.cat  (Corresponding author)}\textsuperscript{1,4} \par \bigskip
  
  \textsuperscript{1}Centre de Recerca Matemàtica (CRM). Campus de Bellaterra. Edifici C, Cerdanyola del Vallès 08193 Barcelona, Spain
  
  \textsuperscript{2}Departament de Matemàtiques. Universitat Politècnica de Catalunya (UPC). Avda. Diagonal 647, 08028 Barcelona, Spain
  
  \textsuperscript{3}Institute of Mathematics of the UPC-BarcelonaTech (IMTech). C. Pau Gargallo 14, 08028 Barcelona, Spain

  \textsuperscript{4}Dynamical Systems and Computational Virology. CSIC Associated Unit CRM-I$^2$SysBio
  
  \textsuperscript{5}Institute of Integrative Systems Biology(I$^2$SysBio). CSIC-Universitat de València, Avda. Catedrático Agustín Escardino 9, Paterna, 46980 València, Spain
  
  \textsuperscript{6}The Santa Fe Institute, 1399 Hyde Park Road, Santa Fe, NM 87501, USA

\today
\end{center}

\begin{abstract}
The dynamics of wild-type (wt) RNA viruses and their defective viral genomes (DVGs) have been extensively studied both experimentally and theoretically. This research has paid special attention to the interference effects of DVGs on wt  accumulation, transmission, disease severity, and induction of immunological responses.  This subject is currently a highly active, and promising, area of research since engineered versions of DVGs have been shown to act as  antiviral agents. However, viral infections involving wt, DVGs and other subviral genetic elements, like viral RNA satellites (satRNAs) have received scarce attention. Satellites are molecular parasites genetically different from the wt virus, which exploit the products of the latter for their own replication in as much as DVGs do, and thus they need to coinfect host cells along with the wt virus to complete their replication cycle. Despite satRNAs are very common, their dynamics co-infecting with wt viruses have been poorly investigated. Here, we analyze a mathematical model describing the initial replication phase of a wt virus producing DVGs and coinfecting with a satRNA. The model, which explicitly considers the viral RNA-dependent RNA polymerase produced by the wt virus, has three different dynamical regimes depending upon the wt replication rate ($\alpha$), the fraction of DVGs produced during replication ($\omega$), and the replication rate of the satRNA ($\beta$): ($i$) full extinction when $\beta > \alpha (1 - \omega)$; ($ii$) a bistable regime with full coexistence governed by a quasi-neutral curve of equilibria and full extinction when $\beta = \alpha (1 - \omega)$; and ($iii$) a scenario of bistability separating full extinction from wt-DVGs coexistence with no satRNA when $\beta < \alpha (1 - \omega)$. The transition from scenarios ($i$) to ($iii$) occurs through the creation and destruction of a quasi-neutral curve of equilibria in a global bifurcation that we name as \textit{quasi-neutral nullcline confluence} (QNC) bifurcation: at the bifurcation value, two nullcline hypersurfaces coincide, giving rise to the curve of equilibria. In agreement with previous research on global bifurcations tied to quasi-neutral manifolds, we have identified numerically and derived analytically scaling laws of the form $\tau \sim |\mu|^{-1}$, being $\tau$ the length of the transients close to the remnant curve and $\mu$ the distance to the bifurcation value.

\end{abstract}

\keywords{{\it keywords}: Bifurcations; Complex systems; Defective interfering genomes; Dynamical systems; Satellite RNAs; Subviral particles}

\setcounter{footnote}{0}
\section{Introduction}

The interaction amongst full-genome wild-type (wt) viruses and the plethora of non-standard viral genomes (nsVGs) that may coexist within the same infected host cell is essential to forecast the outcome of  infections and their impact on virulence. Viruses infect cells and unfold their replication mechanisms to carry out a complete infectious cycle by kidnapping host cells' resources. This property turns them into obligate intracellular parasites for their replication and further propagation~\cite{flint2020principles}. Particularly, RNA viruses are characterized by extremely large population sizes and very fast rates of evolution~\cite{Duffy2008}. This is due to the high mutation rates, since the viral-encoded RNA-dependent RNA polymerase (RdRp) generically lacks of proof-reading mechanism~\cite{Ferrer-Orta2006,Sanjuan2010}. Broadly speaking, nsVGs can be classified into two groups~\cite{GonzalezAparicio&Lopez2024}, (\emph{i}) those that are generated by erroneous replication of the wt genome by the viral RdRp and include hypermutated genomes, deletions, insertions, and different types of reorganized genomes, collectively known as defective viral genomes (DVGs); and (\emph{ii}) other RNA genomes that are not genetically related with the wt virus but coinfect with it and could encode (satellite viruses) or not (satellite RNAs --satRNAs) their own proteins.  In common, all these nsVGs are unable of completing a replication cycle in absence of a wt virus that acts as a helper (HV) that provides all the necessary factors~\cite{vignuzzi}.  Among the diverse types of DVGs, some can disrupt viral replication by hijacking proteins encoded by the wt virus. This specific type of DVGs, called defective interfering particles (DIPs), was first identified by Huang and Baltimore in the 1970s~\cite{Huang1970}. 

Coinfections or superinfections\footnote{Coinfection refers to two different viral types entering the cell at the same time. Superinfection refers to the subsequent infection of a host already infected by another viral type.} of the HV with other subviral genetic elements such as satellite viruses and satRNAs can have a big impact on virus dynamics and modulate symptomatology. Satellites are usually non-related genetically to the HV~\cite{Palukaitis2016} but they depend on it for replication, infection and/or movement, as DIPs do. Satellites can influence viral pathogenicity and accumulation. Unlike satRNAs, satellite viruses contain genetic instructions to, \emph{e.g.}, produce a protein to encapsulate their genomes. Virus satellites and satRNAs are very common in plant infections~\cite{Simon2004,Badar2021}, and some satellites are known to infect unicellular eukaryotic cells~\cite{Schmitt2000}, insects~\cite{Ribiere2010}, and vertebrates~\cite{Krupovic2016}. The implications of satellites in ongoing infections cover a wide range of symptoms: from attenuation, for instance due to a decrease of RNA HV accumulation~\cite{GALON199558}, to symptoms aggravation~\cite{Roux1991}, as it is the case for broad bean mottle virus and turnip crinkle virus~\cite{Simon2004,Badar2021}, or for the hepatitis delta virus (HDV). HDV is a virus satellite that infects together with hepatitis B virus (HBV) using the HBV surface antigen to form enveloped particles capable of cell-to-cell transmission~\cite{GIERSCH2014538}. However, it is able to replicate autonomously in the absence of HBV~\cite{Kuo1991}. HDV coinfection with HBV or superinfection of HBV carriers often results in exacerbation of the underlying HBV hepatitis~\cite{Rizzeto1988}, giving rise to more damaging liver disease and a more rapid evolution towards hepatocellular carcinoma. The prevalence of HDV among HBV carriers is 13.02\%, corresponding to 48–60 million infections globally~\cite{Miao2020}. 

Despite the large number of examples of viral coinfections with satellites, the dynamics of these systems formed by hyperparasites (nsVGs) of parasites (HV) have been poorly studied, specially from a dynamical perspective.
%\textbf{Quasi-neutral curve}
In this article we investigate a mathematical model describing the initial phase of the within-cell replication process of an RNA HV producing DVGs that act as DIPs and coinfecting with a satRNA. Previous research explored a similar system considering these three parasitic agents~\cite{LAZARO2024107987}. Our model, as a difference from the one presented in~\cite{LAZARO2024107987}, explicitly includes the dynamics of the RdRp, whose production depends on the amount of HV and supports the replication of the three RNA molecular species, also introducing further nonlinearities into the system. The viral genome for positive-sense single-stranded RNAs acts as a messenger RNA. Once it enters into the host cell, it is directly translated by the cell ribosomes to produce the RdRp \cite{RAMPERSAD201855}. The studied model shows parameter conditions for which the phase space exhibits a quasi-neutral curve made up of equilibrium points allowing the coexistence of the HV and the two subviral elements. 

In dynamical systems, a quasi-neutral manifold refers to a set of equilibrium points that form a continuous manifold, rather than isolated equilibria. This typically occurs when the system has eigenvalues with zero values associated with the linearization at any point on the manifold, indicating neutral stability along certain directions. Such manifolds typically involve that different initial conditions end up in different equilibrium values. Quasi-neutral manifolds have been previously described in dynamical models for viruses, including asymmetric RNA replication modes~\cite{Sardanyes2018} and epidemiological-like models developed to investigate coronaviruses infections in cell cultures. For this latter system, dynamics were multi-stable and governed by quasi-neutral planes filled with equilibria~\cite{Munoz2025}. Quasi-neutral manifolds have been also found in nonlinear systems describing the dynamics of allele fixation~\cite{Parsons2007,Parsons2008}, in models of sexual diploid populations~\cite{Heisler1990,Greenspoon2009}, in host-parasite systems~\cite{Fontich2022}, in a two-species Lotka-Volterra model~\cite{Lin2012}, and in predator-prey dynamical systems displaying quasi-neutral surfaces~\cite{Farkas1984}. Quasi-neutral curves of equilibria are invariant and uniparametric mathematical objects consisting of a continuum of equilibrium points with local attracting or repelling directions together with neutral ones~\cite{wiggins}. They can play an essential role in organizing the global dynamics of a dynamical system to the extent that they stand for the boundary between two qualitatively different regimes. That is to say, basins of attraction of different equilibrium points may change drastically in a neighbourhood of the parameter value for which the quasi-neutral curve exists as well as the topological structure of the phase portrait~\cite{Fontich2022}. Here, we describe a new global bifurcation tied to this quasi-neutral curve that we have named as \textit{quasi-neutral nullcline confluence} (\QNC) bifurcation. This bifurcation involves the creation and destruction of a curve of equilibrium points which arises for specific parameter conditions making two three-dimensional nullclines to coincide. Similar global bifurcations have been previously described in lower dimensions \cite{Fontich2022}. Our investigation focuses on the dynamics in a neighbourhood of these parameter conditions, in which nearby orbits passing close to a remnant of the (non-existing) quasi-neutral curve remain for a long time close to it, experiencing a kind of ghost effect which largely delays the orbits. Ghosts and bottlenecks causing a slowing-down of dynamics have been widely studied for local bifurcations~\cite{Strogatz1989,Sardanyes2006,Sardanyes2007,Sardanyes2020}. More recently, this concept has been generalized by the so-called ghost channels and ghost cycles. The implications of these ghost structures have been discussed in the context of stability and stochastic dynamics~\cite{koch2024}.

The manuscript is organized as follows. In Section ~\ref{math_model} we introduce the mathematical model describing the dynamics of complementation and competition of a tripartite system composed by a wt virus producing DIPs under the presence of a satRNA. Section~\ref{Results} contains the investigation of the dynamics' domain, nullclines, and invariant sets. Here, the equilibrium points and some global aspects of the dynamics are inspected. Then, we focus on the geometry and the dynamics tied to the quasi-neutral curve of equilibria. We also analyze a global bifurcation involving the disappearance of this quasi-neutral curve and we describe a new mechanism of slowing down tied to the remnant of this curve (a kind of ghost channel). This slowing down is shown to be governed by scaling laws with exponent $-1$ with respect to a close distance between the bifurcation parameter before and after the bifurcation value. Some conclusions about the possible roles of these dynamics in early viral infections are finally outlined in Section~\ref{discussion}. In order to ease the reading, the proofs of all the relevant results presented in all sections are provided in Appendices~\ref{se:appendix:1} and \ref{se:appendix:2}. In some cases, if the proof is straightforward, it has been omitted.

\begin{figure}
\captionsetup{width=\linewidth}
{\includegraphics[width=\textwidth]{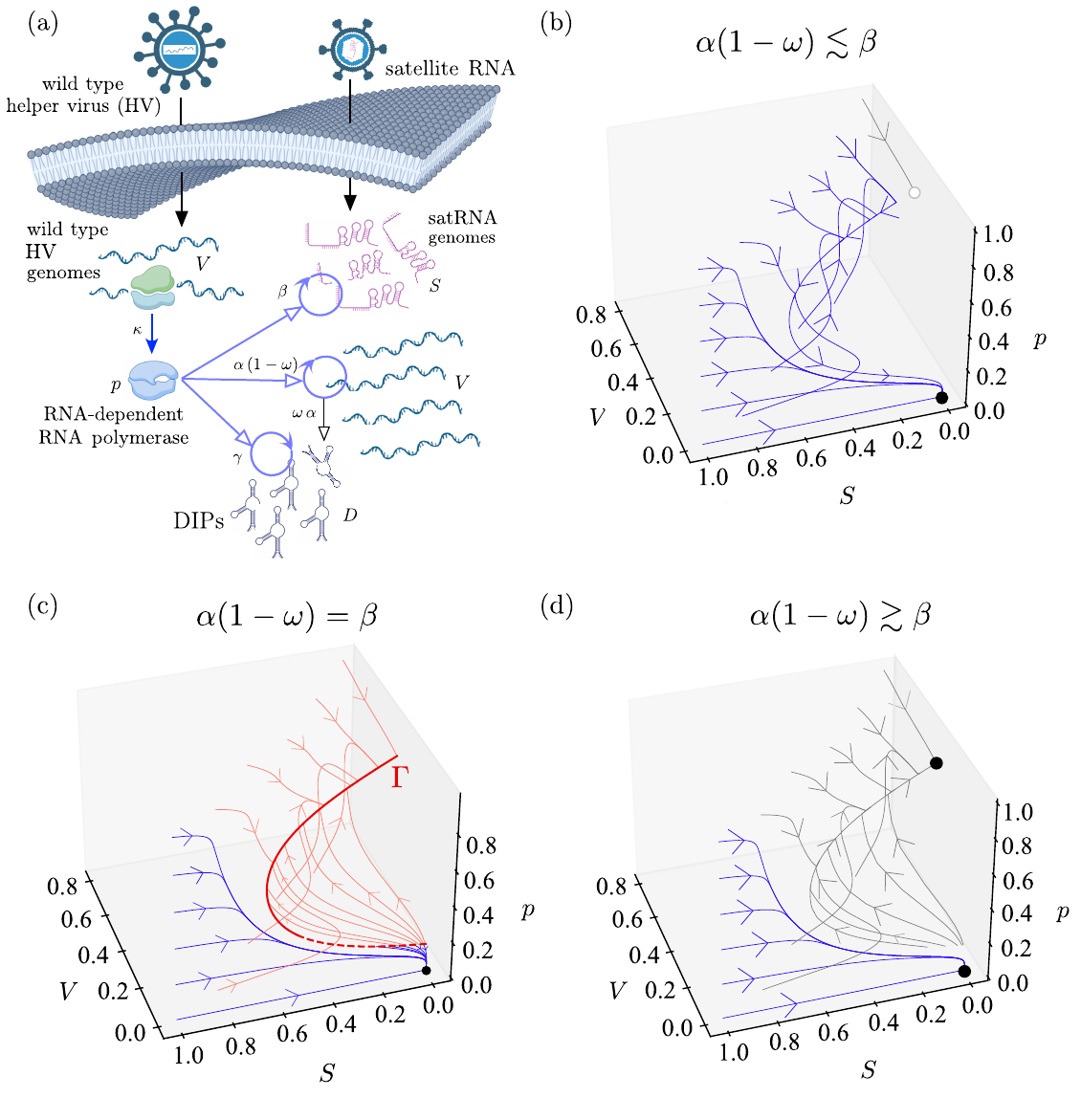}}
    \caption{\small (a) Schematic diagram of the main interactions modeled for the tripartite system wild-type helper virus (HV), its interfering defective particles (DIPs), and a satellite RNA (satRNA) coinfecting the same cell. The HV RNA genome is translated into the viral RNA-dependent RNA polymerase (RdRp). The RdRp replicates the HV  genomes producing DIPs at a rate $\omega$. The RdRp also supports the replication of the DIPs and the satRNA. It is assumed that all the viral agents compete for cellular resources such as nucleotides. Panels (b-d) display phase portraits for three qualitatively different scenarios which depend on the effective replication rate of the HV $\alpha (1-\omega)$ and the replication rate of the satRNA $\beta$. (c) Quasi-neutral curve $\Gamma$ (red curve with locally stable (solid line) and unstable (dashed line) branches). Here, black dots denote local asymptotically stable equilibrium points. White dots are equilibrium points of saddle type with three-dimensional stable manifolds and an unstable one. Blue orbits belong to the basin of attraction of the origin $Q_0$; grey orbits to the basin of attraction of the satRNA-extinction equilibrium point $Q_1$, and the red ones to the basin of attraction of coexistence equilibria $Q_2$.}
    \label{fig_processes}
\end{figure}

\section{Mathematical model}\label{math_model}

Let us define $x = (V,S,D)$ as the vector of state variables containing the population densities of the three RNA species: the helper virus ($V$), the defective interfering genomes ($D$), and the satellite RNA ($S$). Let us also consider the viral RNA-dependent RNA polymerase ($p$), encoded by the HV and supporting the replication of the three RNA species, as another state variable. The within-cell dynamics for this system [schematized in Fig.~\ref{fig_processes}(a)] can be described by the following system of autonomous ordinary differential equations:
\begin{eqnarray}
    \dot{V} &=& \alpha \, (1 - \omega) \,V \,p \, \Omega(x) - \eps \,V, \label{eq1} \\
    \dot{S} &=& \beta \, S\, p \, \Omega(x) - \eps \, S,\label{eq4} \\
    \dot{D} &=& \left(  \omega \, \alpha \, V  + \gamma  \, D \right)\, p \,  \Omega(x) - \eps \, D, \label{eq5}\\
    \dot{p} &=& \kappa \, V (1 - p) - \varepsilon_p \, p,\label{eq2}
\end{eqnarray}
with $\Omega(x) = 1 - V - S- D$. The model considers the processes of RdRp translation, genome replication, production of DVGs, competition, complementation, and RNA degradation. Concerning the viral RdRp, its dynamic equation describes the processes of enzyme production and degradation. The logistic-like term $\Omega(x)$, assuming normalized carrying capacity, introduces intra- and inter-specific competition between the viral agents. Such a competition is assumed to occur due to cellular finite resources i.e., mononucleotides. A different logistic function is used for the synthesis of the RdRp. For simplicity, we focus on the initial replication stages without modeling the encapsidation and formation of virions. 

The model parameters are given by the replication rates of the HV ($\alpha > 0$), the satRNA ($\beta > 0$), and the DIPs ($\gamma > 0$). Despite the replication rates for the DIPs and the satRNA may be expected to be faster than the one of the HV, mainly due to the differences in genome  length (see Tables $1-5$ in~\cite{Badar2021} for genome lengths of satRNAs), we will provide a general investigation of these parameters. The replication speed of each of the RNA species will certainly depend on the length of their genomes but other factors can also influence on these relative replication rates. These can include the secondary structures of viral RNAs \emph{e.g.}, 
stem-loops can act as physical barriers slowing down the RdRp advance over the RNA template molecule and require unwinding~\cite{Thorne2014}. Related to this unwinding process, the availability of helicases can also be of importance in facilitating replication~\cite{Raney2010}, as well as nucleotides' availability~\cite{Takeda2018}. Moreover, host proteins such as hnRNPs can influence replication speed either facilitating or inhibiting replication~\cite{Kim2020}, or the capability of RNAs to access the replication factories associated to the endoplasmic reticulum~\cite{Romero2014}. Moreover, the investigation of the dynamics for arbitrary parameter values within biologically-meaningful domains may allow for the possible adaptation of their values for particular viruses, especially concerning the synthesis rates of new genomes. The model also considers that DIPs are produced at a rate $\omega \in (0, 1]$ due to errors during the process of HV transcription. In addition, we consider that the RdRp is produced at a constant rate $\kappa>0$ and proportionally to the amount of HV genomes. Finally, the three RNA species are assumed to degrade at a constant rate $\varepsilon >0$ while RdRp degrades at rate $\varepsilon_p > 0$ [see Fig.~\ref{fig_processes}(a)] . 

As we discuss below, the different dynamics of the modeled dynamical system have a strong dependence on the effective HV replication given by $\alpha (1-\omega)$, and on the replication rate of the satRNA $\beta$. This fact is illustrated in Fig.~\ref{fig_processes}(b-d).

\section{Results}\label{Results}
We first carry out an analytical description of the vector field described by Eqs.~\eqref{eq1}-\eqref{eq2}. To do so, we define the biologically meaningful domain of dynamics, the expression for the nullclines and some invariant subsets. Then, we present two main results outlining different conditions for the existence of equilibrium points different from the origin. 

\subsection{Dynamics domain, nullclines and invariant subsets}\label{dd:n:is}
According to former considerations, the dynamics of our system is confined to the following set in $\R^4$:
\begin{equation}\label{domain}
    \mathcal{U} = \bigg\{(V,S,D,p) \in \mathbb{R}^4 \quad \Big| \quad V, D, S \geq 0; \quad 0 \leq p \leq 1 \quad \text{and} \quad V + D + S \leq 1\bigg\}.
\end{equation}
Indeed, $\mathcal{U}$ is positively invariant since the hyperplanes $V=0$ and $S=0$ are invariant subsets; when $p = 0$, we have $\dot{p} > 0$ and when $p = 1$, we have $\dot{p} < 0$; for $D=0$, $\dot{D} > 0$ and at $\Omega(x) = 0$, $\dot{V} <0$, $\dot{D} < 0$ and $\dot{S} < 0$ in such a way that the flow points inwards the domain from all its boundaries. In fact, $\mathcal{U}$ consists of the Cartesian product between a $3$-dimensional tetrahedron spanned by the three RNA populations, with finite faces at the intersection of planes $V=0$, $S=0$, $D=0$ and $V + S + D = 1$; and the finite and closed real interval $[0,1]$, which is the domain of the state variable $p$.

The $V$-nullcline has two components given by the invariant set $V = 0$ and the hyperquadric with expression
\begin{equation}\label{nullV}
    H_1 : p\,\Omega(x) = \frac{\varepsilon}{\alpha (1-\omega)} \Longleftrightarrow p(1 - V - D - S) = \frac{\varepsilon}{\alpha (1-\omega)}.
\end{equation}
Nullcline $\dot{S} = 0$ also exhibits two components given by the invariant set $S=0$ and the hyperquadric embedded in the 4th-dimensional space with the following expression:
\begin{equation}\label{nullS}
    H_2 : p\,\Omega(x) = \frac{\varepsilon}{\beta} \Longleftrightarrow p (1-V-D-S) = \frac{\varepsilon}{\beta}.
\end{equation}
Dynamics in the invariant hyperplane $S=0$ do not modify the general dynamics since the satRNA  only depends on the RdRp to persist excluding competition processes. The nullcline $\dot{D} = 0$ is the algebraic surface $(\omega\alpha V + \gamma D) p \Omega(x) = \varepsilon D$ and nullcline $\dot{p} = 0$ is provided in the lemma below.
\begin{lemma}\label{p_equilibrium}
    Let $Q^*=(V^*, S^*, D^*, p^*)$ be any equilibrium point of system~\eqref{eq1}-\eqref{eq2}. Then,
    \begin{equation}
        p^*=p^*(V^*)=1 - \frac{\eps_p}{\kappa V^* + \eps_p},
    \label{pequilib}    
    \end{equation}
    which corresponds to nullcline $\dot{p} = 0$. It satisfies $0<p^* <1$ and, moreover,  $p^*=0$ if and only if $V^*=0$. 
\end{lemma}

Nullclines \eqref{nullV}-\eqref{nullS} are parallel hypersurfaces which only intersect when they coincide for $\alpha(1-\omega) = \beta$. In order for both of them intersecting $\mathcal{U}$, conditions $\varepsilon \leq \alpha(1-\omega)$ and $\varepsilon \leq \beta$ must be fulfilled, which are reasonable restrictions since degradation rates are assumed to be much lower than the effective replication rates\footnote{By effective replication rate we mean the net replication rate of the HV, given by $\alpha(1-\omega)$. That is, the fraction of the HV that replicates successfully with no mutations resulting in additional DIPs.} for viral particles. Previous quantitative research on positive-sense single-stranded RNA viruses has shown that genomes' replication rates are about one or two orders of magnitude larger than their degradation rates~\cite{Martinez2011}. When both $H_1$ and $H_2$ intersect $\mathcal{U}$, and given their parallelism, the 4-dimensional space is divided into three different disjoint sets. Assuming $\varepsilon \neq 0$, the intersection of $H_1$ and $H_2$ with $p=0$ and $1 = V + D + S$ is empty and only the intersection with $V=0$, $S=0$, $D=0$ and $p=1$ is non-empty.

Subsets $V=0$ and $S=0$ are invariant. Absence of $V$ leads to a completely unstable configuration  from a biological point of view since $\dot{p} = -\varepsilon_p p$ makes RdRp density to strictly decrease for hypothetical initial conditions satisfying $p_0 \neq 0$ and thus, $D$ and $S$ populations depending on it would decay at certain point driving the system to total extinction. Therefore, in general, viral replication is prone to progress subject to the capacity of the HV to synthesize RdRp. On the other hand, dynamics within $S=0$ are governed by the existence of $V$-$D$-$p$ coexistence equilibrium points. Invariability of $S=0$ makes dynamics in the absence of satRNA to be completely independent to those dynamics with the 4 variables. This behaviour is biologically expected since satRNAs are parasites of the HV\footnote{satRNAs are molecular hyperparasites of the HV, which is also a parasite.} and, therefore, removing them just results in fewer viral agents taking profit of the host cell resources (less competition) and the RdRp synthesized by the HV.

\subsection{Equilibrium points and non-existence of periodic orbits}
\label{se:equilibrium:points}
This section is devoted to describe the invariant sets for Eqs.~\eqref{eq1}-\eqref{eq2}. Specifically, it deals with the existence of equilibrium points and periodic orbits. The first result rules out the existence of periodic orbits and, consequently, the existence of limit cycles.
\begin{prop}[Non-existence of periodic orbits] \label{PO}
    There are no $T$-periodic solutions of system (\ref{eq1})-(\ref{eq2}) for any $T>0$ with initial conditions in $\mathcal{U}$.
\end{prop}

An analysis of the equilibrium points of system \eqref{eq1}-\eqref{eq2} gives rise to the results below. 
First, two different conditions are given ensuring an equilibrium point to be the origin, \emph{i.e.}, the populations' full extinction.

\begin{lemma}
    Let $Q^*=(V^*, S^*, D^*, p^*)$ be an equilibrium point of system~\eqref{eq1}-\eqref{eq2} and $V^*=0$. Then, necessarily, $Q^*$ is the origin, i.e. $Q^*=Q_0=(0,0,0,0)$.    
    \label{eq:V:zero}    
\end{lemma} 

\begin{lemma}\label{gamma:alpha:Q0}
    Let $Q^*=(V^*,S^*,D^*,p^*)$ be an equilibrium point of system~\eqref{eq1}-\eqref{eq2} and $\alpha (1-\omega) = \gamma$. Then, necessarily, $Q^*$ is the origin, i.e. $Q^* = Q_0 = (0,0,0,0)$.
\end{lemma}

Hence, by virtue of Lemma~\ref{gamma:alpha:Q0}, when the wt virus $V$ and its DIPs $D$ replicate at exactly the same effective rate, the only possible equilibrium point is the origin. The local stability for the origin is provided in the following result for all values of the parameters.
\begin{lemma}\label{Q0:local:stability}
    The origin $Q_0$ is always locally asymptotically stable for any parameter values.
\end{lemma}
Let $Q^* = (V^*,S^*,D^*,p^*)$ be an equilibrium point of system \eqref{eq1}-\eqref{eq2}. Let us assume now that $V^*\ne 0$ and seek for other equilibrium points different from the origin. In particular, from equation $\dot{S}=0$ and assuming $\beta \ne 0$ one has
\begin{equation}
    S^*\beta \left( p \Omega(x) - \frac{\eps}{\beta} \right) = 0,
\end{equation}
giving rise to two different cases: equilibrium points such that $S^* = 0$, denoted by $Q_1$ (satRNA extinction equilibrium points); and equilibrium points such that $S^* \neq 0$, denoted by $Q_2$ (coexistence equilibrium points). Propositions below identify the different parameter conditions that must hold for
$Q_1$-equilibrium points (Proposition~\ref{Q1_prop}) and 
$Q_2$-equilibrium points (Proposition~\ref{Q2_prop}) to exist.

\begin{prop}[SatRNA extinction equilibrium points]\label{Q1_prop}
    Let us denote by $Q_1 = (V_1, 0, D_1, p_1)$ a satellite-free equilibrium point of system~\eqref{eq1}-\eqref{eq2} (\emph{i.e.}, with $S_1 = 0$). Then, a necessary condition for $Q_1$ to exist is that
    \begin{equation}
    \alpha (1-\omega) > \max\{ \gamma, \eps\}.
    \label{eq:Q1:necessary:cond}
    \end{equation}
    Let assume that condition~\eqref{eq:Q1:necessary:cond} holds and let us define
    the quadratic polynomial $q(V) = T_2 V^2 + T_1 V + T_0$, with coefficients
    \begin{equation*}
        T_2 = \frac{\varepsilon \kappa (\alpha - \gamma)}{\alpha (1-\omega) - \gamma}, \qquad T_1 = - \kappa \varepsilon \left(1 - \frac{\varepsilon}{\alpha(1-\omega)}\right), \qquad T_0 = \frac{\varepsilon_p \varepsilon^2}{\alpha (1-\omega)},
    \end{equation*}
    and discriminant $\Delta = T_1^2 - 4T_0 T_2$. Then, there are at most two points $Q_1=(V_1,0,D_1,p_1) \in \mathcal{U}$, given by 
    \begin{equation}
        q(V_1)=0, \qquad 
        p_1(V_1) = 1 -   \frac{\varepsilon_p}{\kappa V_1 + \varepsilon_p}, \qquad D_1(V_1) = \frac{\alpha \omega}{\alpha(1-\omega) - \gamma} V_1
    \end{equation}
    which are equilibrium points of the system~\eqref{eq1}-\eqref{eq2}. These two points coincide when $\Delta$ vanishes.
\end{prop}

\begin{remark}
    From a biological viewpoint, condition \eqref{eq:Q1:necessary:cond} corresponds to the fact that the effective growth rate of the HV is larger than both its own degradation and the DIP intrinsic growth rate.      
\end{remark}

\begin{prop}[Coexistence equilibrium points]\label{Q2_prop}
    Let $Q_2 = (V_2,S_2,D_2,p_2)$ be an equilibrium point of system~\eqref{eq1}-\eqref{eq2} with $S_2 \neq 0$. From Lemma~\ref{eq:V:zero} we know that $V_2\neq 0$.
    Then, necessarily, we must have
    \begin{equation}
    \alpha (1 - \omega) = \beta 
    \qquad \textrm{and} \qquad \beta > \gamma.
    \label{cond:eq:coex}
    \end{equation}
    Let us assume that conditions~\eqref{cond:eq:coex} hold. Thus, 
    $Q_2 = (V_2,S_2,D_2,p_2)$ is a coexistence equilibrium point if it satisfies that
    \begin{equation}
    p_2 = p_2(V_2) = 1 - \frac{\varepsilon_p}{\kappa V_2 + \varepsilon_p}, \qquad D_2 = \frac{\beta \omega}{(\beta-\gamma)(1-\omega)} \, V_2,
    \label{eq:p2:D2}
    %\label{eq:D2}
    \end{equation}
    and $(V_2,S_2)$ belongs to the piece of the conic 
    \begin{equation}\label{pol_q2}
        q_2(V_2, S_2) = -\kappa \left(1 + \frac{\alpha \omega}{\beta - \gamma}\right) V_2^2 + \kappa \left(1 - S_2 - \frac{\varepsilon}{\beta}\right) V_2 - \frac{\varepsilon \varepsilon_p}{\beta} =0
    \end{equation}
   which falls in $\mathring{\mathcal{U}}$.
\end{prop} 
The polynomial $q(V)$ in Proposition~\ref{Q1_prop} is a multiple of the polynomial $q_2(V_2,0)$ given by \eqref{pol_q2}, as expected, if we take into account that $\alpha(1-\omega) = \beta$ in $q_2(V_2,S_2)$. Precisely, $q(V) = -\varepsilon q_2(V_2,0)$.
Furthermore, if $S_2$ satisfies that
\begin{equation}
        2\kappa \left(1 + \frac{\alpha \omega}{\beta - \gamma}\right)V_2 \neq \kappa \left(1 - S_2 - \frac{\varepsilon}{\beta}\right),
    \end{equation}
then we have that $V_2=f(S_2)$ verifies $q_2(f(S_2),S_2)=0$ where
    \begin{equation}\label{fS_numerical}
        f(S_2) = \frac{-\kappa \left(1 - S_2 - \frac{\varepsilon}{\beta}\right) \pm \sqrt{\kappa^2 \left(1 - S_2 - \frac{\varepsilon}{\beta}\right)^2 - 4 \kappa \left(1 + \frac{\alpha \omega}{\beta - \gamma}\right) \frac{\varepsilon\varepsilon_p}{\beta}}}{-2\kappa \left(1 + \frac{\alpha \omega}{\beta - \gamma}\right)}.
    \end{equation}
This explicit expression has been employed throughout this work in the numerical computation of the $Q_2$-equilibrium points.

Notice that conditions in~\eqref{cond:eq:coex} can be biologically interpreted as: $(a)$ $\alpha(1-\omega)>\gamma$, the HV's effective growth rate is larger than the DIPs' intrinsic replication rate; and ($b$) $\beta>\gamma$, that is, both the HV and the satRNAs must replicate at a higher effective rate than the DIPs.    

The existence of equilibrium points $Q_1$ and $Q_2$ is then a consequence of the balance between the effective replication rate of the HV ($\alpha(1-\omega)$), the intrinsic replication rates of both the satRNA ($\beta$) and the DIPs ($\gamma$), and the degradation rates of the viral populations ($\varepsilon$). The satRNA extinction equilibrium $Q_1$ requires that HV replicates faster than the DIPs and faster than its own degradation. On the other hand, coexistence equilibria are possible only provided that the HV and satRNAs replicate at exactly the same pace and faster than DIPs. 

\subsection{Geometry and dynamics of the coexistence quasi-neutral curve}\label{gamma_geometry}

Let us denote by $\Gamma$ the curve of equilibrium points $Q_2$ provided by Proposition~\ref{Q2_prop}. If nothing is explicitly said, we will
reduce the study of such $\Gamma$ to its projection on the space $(V,S,D)$. Moreover, along this section, conditions~\eqref{cond:eq:coex} and $V\ne 0$ will be always assumed and, for the sake of simplicity, we will refer to the flow induced by the system \eqref{eq1}-\eqref{eq2} simply as \emph{the flow}.

\begin{figure}
    \centering
    \captionsetup{width=\linewidth}
    \includegraphics[width=\linewidth]{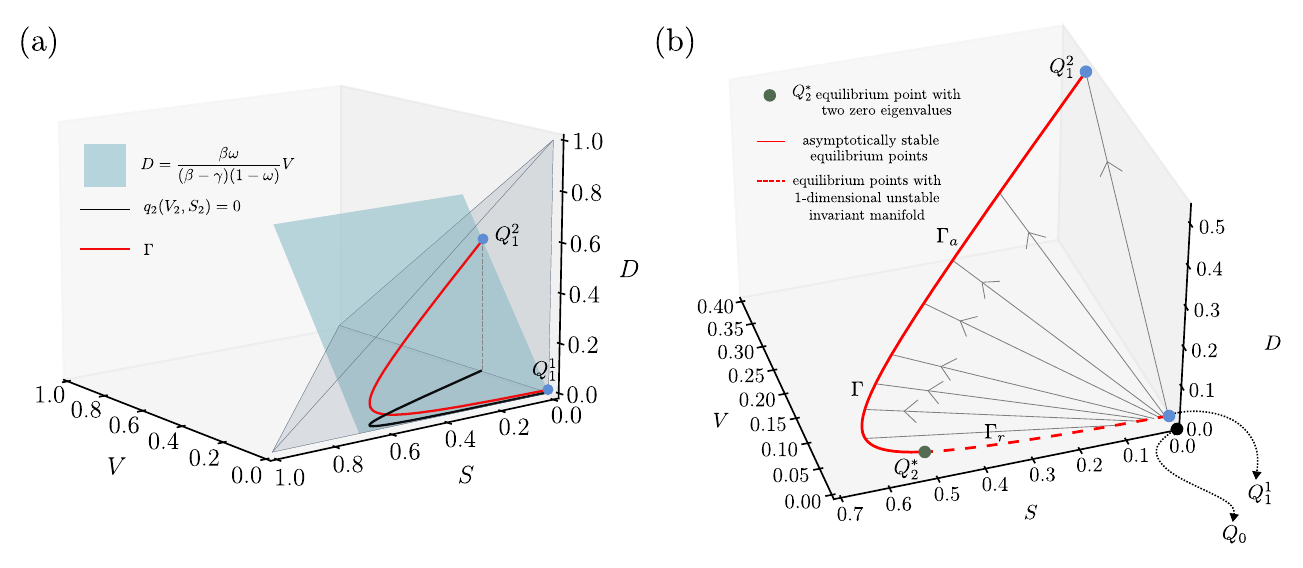}
    \caption{\small (a) Projection of the quasi-neutral curve $\Gamma$ (red) onto the space $(V,S,D)$. $\Gamma$ is contained in the invariant plane $\Pi_{DV}$ (blue surface) and embedded into the projection of $\mathcal{U}$ onto $(V,S,D)$ (grey tetrahedron). (b) Schematic projection in the space $(V,S,D)$ of the heteroclinic connections among $Q_2$-points in $\Gamma$. Notice that, in particular, these connections belong to the homoclinic invariant manifold of the curve $\Gamma$ itself.
    The heteroclinic connections (straight lines) go in infinite time from the unstable equilibrium points in $\Gamma_r$ (dashed red curve) to the locally attracting equilibrium points of the piece of curve $\Gamma_a$ (in solid red color). 
    The dark-green point in $\Gamma$ separates both branches, being tangent to the plane $S/V=c^*$ and having two zero eigenvalues in its jacobian matrix. The equilibrium points $Q_1^1$ and $Q_1^2$ (in light blue color) represent the intersection of $\Gamma$ with the invariant plane $\{ S=0\}$.} 
    \label{QNC_Gamma}
\end{figure}

From its definition it is clear that $\Gamma$ is a piece of hyperbola in $\mathcal{U}$ contained in the plane
\begin{equation}\label{D:hyperplane}
    \Pi_{DV} : \quad D = \frac{\beta\omega}{(\beta - \gamma)(1-\omega)} V
\end{equation}
(see Fig.~\ref{QNC_Gamma}(a) for a geometrical representation).
The following two propositions determine
crucial constraints for the dynamics of this case and which do not hold in the general parameter situation.
\begin{prop}\label{D:invariant:surface}
The plane $\Pi_{DV}$ is invariant by the flow.
\end{prop}
\begin{prop}\label{first:integral:prop}The function $S/V$ is a first integral of system~\eqref{eq1}-\eqref{eq2}, that is, 
\begin{equation*}
        \frac{S(t)}{V(t)} = \frac{S(0)}{V(0)} \qquad \forall t \geq 0.
    \end{equation*}
\end{prop}
\begin{minipage}[t]{0.32\textwidth}
  \raggedright
  \raisebox{\dimexpr \topskip-\height}{%
  \includegraphics[width=\textwidth]{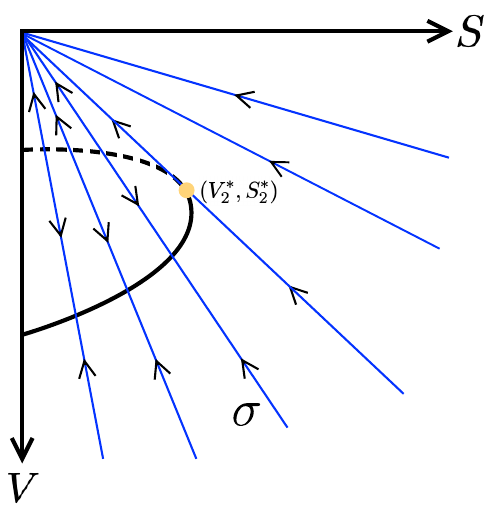}  }
  \captionsetup{width=0.94\linewidth}
  \captionof{figure}{Projection of solutions of Eqs.~\eqref{eq1}-\eqref{eq2} onto the plane $(V,S)$ for $\alpha(1-\omega) = \beta$. The arrows indicate the direction of the orbits in positive time.}
  \label{orbits_VS}
\end{minipage}\hfill
\begin{minipage}[t]{0.6\textwidth}
The existence of the first integral $S/V$, together with the invariance of $\Pi_{DV}$, provides information on the geometry of the orbits when $\alpha(1-\omega) = \beta$. Indeed, in the $(V,S,D)$-space, any straight line $\sigma$ which is intersection of the two planes $S/V=c$ and $\Pi_{VD}$ is also invariant by the flow: given any initial condition on $\sigma$, the corresponding orbit stays in $\sigma$ for $t\in \R$ and, actually, $\sigma$ itself is an orbit of the system. Since $\Gamma$ is a branch of a hyperbola contained in $\Pi_{DV}$, any line $\sigma$ must intersect it in two points, one point or none. Let assume that it does so in two points $\{ Q_2^r, Q_2^a\}$\footnotemark
\text{ }which, recall, are equilibrium points. Then $\sigma$ is the orbit of system~\eqref{eq1}-\eqref{eq2} connecting these two points (in infinite time) in a so-called heteroclinic connection. Thus, one of the equilibrium points, say $Q_2^r$, must have a $1$-dimensional unstable invariant curve which coincides with $\sigma$ and the other one, $Q_2^a$, must have a $1$-dimensional stable invariant curve which matches $\sigma$. Notice that, in particular, these point-to-point in $\Gamma$ heteroclinic connections fall within the more general homoclinic invariant manifold of the curve $\Gamma$ to itself. 
\end{minipage}
\footnotetext{Superscripts $r$ and $a$ stand for, respectively, repeller and attractor.}

Consequently, the curve $\Gamma$ splits into an stable branch $\Gamma_a$ [red solid curve in Figure~\ref{QNC_Gamma}(b)] and a unstable branch $\Gamma_r$ [red dashed curve in Figure~\ref{QNC_Gamma}(b)], both joined by orbits (in gray) whose projection on the $(V,S,D)$-space are straight lines. The point $Q_2^*\in \Gamma$ separating these two branches [dark-green dot in Figure~\ref{QNC_Gamma}(b)] belongs to the straight line $\sigma$ in $S/V=c^*$, contained in $\Pi_{DV}$, which is tangent to $\Gamma$. The projection of such orbits on the plane $(V,S)$ is displayed in Fig.~\ref{orbits_VS}.

\begin{figure}
    \centering
    \captionsetup{width = \linewidth}
    \includegraphics[width=0.87\textwidth]{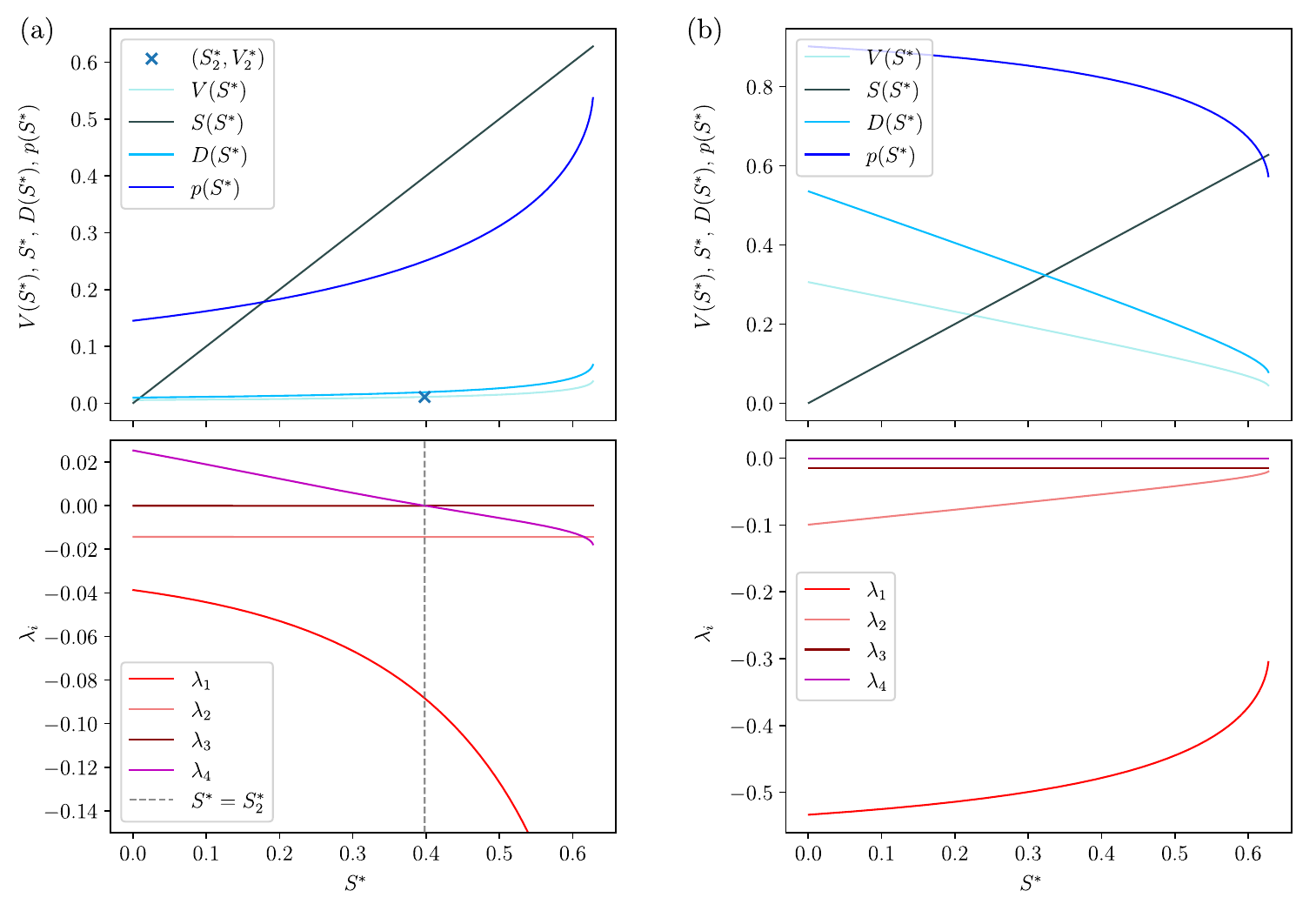}
    \caption{\small Coordinates of equilibrium points $Q_2$ filling $\Gamma$ (upper panels) and the corresponding eigenvalues of their differential matrix (lower panels). The parameter values used satisfy conditions in~\eqref{cond:eq:coex} and are given by $\alpha=0.875$, $\beta = 0.7$, $\omega = 0.2$, $\gamma = 0.6$, $\kappa = 0.3$, $\varepsilon = 0.1$ and $\varepsilon_p = 0.01$. Panels (a) and (b) are related, respectively, to the two possible expressions for $V_2$ given by~\eqref{fS_numerical}. The point $(S_2^*,V_2^*)$,    marked with a cross, is given by expression~\eqref{double:0:eigval:coord} and corresponds to the $(V,S)$-coordinates of $Q_2^*$.}
    \label{fig:vaps:corba:gamma}
\end{figure}

The local stability of the equilibrium points $Q_2 \in \Gamma$ is determined by the eigenvalues of the differential matrix of system \eqref{eq1}-\eqref{eq2} when evaluated at $Q_2$. In its general form, it is given by:
\begin{equation}\label{dif_matrix_Q2}
    DF(Q_2) = \left(
    \begin{matrix}
        -\beta V_2 p_2 & -\beta V_2 p_2 & -\beta V_2 p_2 & \varepsilon V_2/p_2 \\
        -\beta S_2 p_2 & -\beta S_2 p_2 & -\beta S_2 p_2 & \varepsilon S_2/p_2 \\
        A_1 & A_2 & A_3 & A_4 \\
        \kappa (1-p_2) & 0 & 0 & -\kappa V_2 - \varepsilon_p
    \end{matrix}
    \right),
\end{equation}
where
\begin{align*}
    A_1 &= \frac{\varepsilon (\alpha - \beta)}{\beta} - p_2 (V_2 (\alpha - \beta) + \gamma D_2), \\
    A_2 &= - p_2 (V_2 (\alpha - \beta) + \gamma D_2), \\
    A_3 &= \frac{\varepsilon \gamma}{\beta} - p_2 (V_2 (\alpha - \beta) + \gamma D_2) - \varepsilon, \\
    A_4 &= \frac{\varepsilon}{p_2 \beta} (V_2(\alpha - \beta) + \gamma D_2).
\end{align*}
Recall that the values of $V_2, D_2, p_2$ of $Q_2\in \Gamma$ are given, respectively, by the roots of~\eqref{pol_q2} and expressions in~\eqref{eq:p2:D2}, for $S_2 \in (0,1)$.
For any $Q_2$, the corresponding jacobian matrix~\eqref{dif_matrix_Q2} has a zero eigenvalue (a neutral direction) which corresponds to the tangent direction to $\Gamma$ contained in $\Pi_{DV}$. The stable [$\Gamma_a$, continuous red curve in Fig.~\ref{QNC_Gamma}(b)] and unstable [$\Gamma_r$, dashed red curve in Fig.~\ref{QNC_Gamma}(b)] branches constituting $\Gamma$ join at the point $Q_2^* \in \{S/V = c^*\} \cap \Pi_{DV}$. The value of $c^*$ can be analytically computed and is given by
\begin{equation}
    c^* = \frac{\kappa \beta}{4\varepsilon\varepsilon_p} \left(1 - \frac{\varepsilon}{\beta}\right)^2 - 1 - \frac{\alpha\omega}{\beta - \gamma},
\end{equation}
and its corresponding $V_2$ and $S_2$ coordinates are: 
\begin{equation}\label{double:0:eigval:coord}
    V_2^* = \frac{2\varepsilon\varepsilon_p}{\kappa (\beta- \varepsilon)}, \qquad S_2^* = \frac{2\varepsilon\varepsilon_p}{\kappa (\beta- \varepsilon)} \left(\frac{\kappa \beta}{4\varepsilon\varepsilon_p} \left(1 - \frac{\varepsilon}{\beta}\right)^2 - 1 - \frac{\alpha\omega}{\beta - \gamma}\right).
\end{equation} 
In addition to this common zero eigenvalue (associated to $\Gamma$),
the differential matrix $DF(Q_2)$ for $Q_2\in \Gamma_a$ 
has three real negative eigenvalues [two in $(V,S,D)$], and so it is of attracting type. On the other hand, any point $Q_2\in \Gamma_r$ has
three real eigenvalues, two negative and one positive. So, it is of saddle-type, with a $1$-dimensional unstable invariant curve.
Regarding the junction point $Q_2^*$, its differential matrix has an extra zero eigenvalue. In some sense this fact can be seen as a kind of transcritical bifurcation \emph{inside $\Gamma$}: one of the real eigenvalues of the equilibrium points filling $\Gamma_r$ changes its sign when it crosses $Q_2^*$, moving along $\Gamma$ in the direction of $\Gamma_a$. The location of the point $Q_2^*$ in $\Gamma$ depends on the rest of the parameters of the system. This is shown in Fig.~\ref{fig:vaps:corba:gamma}(a), which displays the coordinates of equilibrium points $Q_2$ in $\Gamma$ as a function of $S^*$. All these eigenvalues have been numerically computed during the simulations.

This particular type of invariant curves (like $\Gamma$), formed by equilibrium points, has been usually referred in the literature as \textit{quasi-neutral} curves~\cite{arderiu,Munoz2025,Parsons2007,Parsons2008,Heisler1990,Greenspoon2009,Fontich2022,Lin2012,Farkas1984}.

\subsection{Global bifurcation associated to the breakdown of $\Gamma$}
As seen in Section~\ref{se:equilibrium:points}, three different kind of equilibrium points may exist depending on the parameter values: the origin $Q_0$ as a coextinction state, the satellite-free state $Q_1$, and the coexistence of the three viral populations and the RdRp, $Q_2$. The latter, which exists only under specific conditions, leads to the following definition.
\begin{defi}
   The \emph{critical replication rate of HV}, denoted by $\alpha^*$, is the value of $\alpha$ satisfying
    \begin{equation}
        \alpha^* (1-\omega) = \beta.
    \end{equation}
Thus, if $\alpha$ is taken as a bifurcation parameter we denote by
    \begin{equation}
        \mu = \alpha - \alpha^*
    \end{equation}
    the distance to this critical value. It satisfies $\mu > 0$ for $\alpha > \alpha^*$ and $\mu < 0$ for $\alpha < \alpha^*$.
\end{defi}
The critical rate $\alpha^*$ corresponds to the precise value $\alpha$ (provided $\beta>\gamma$ also holds) at which the curve $\Gamma$ of coexistence equilibrium points $Q_2$ exists, already analised in Section~\ref{gamma_geometry}.
From a biological viewpoint, this case implies the equality between the effective replication rate of HV and the satRNA's replication rate.

The aim of this section is to study this global bifurcation - that we have named \emph{quasi-neutral nullcline confluence} (\QNC) bifurcation - 
that system~\eqref{eq1}-\eqref{eq2} undergoes at $\mu = 0$ (for $\beta > \gamma$). As far as we know, this type of bifurcation has not been previously described. This change in the dynamics is linked to an abrupt transition from a monostability to a bistability global scenarios. The relevance of this case is that such bifurcation shows up through the exceptional appearance and breakdown of the quasi-neutral curve of coexistence equilibria $\Gamma$. Even more, the locally attractive property of this curve for $\mu=0$ will make, for values of $\mu \sim 0$, that nearby orbits (a remarkable domain in the $(V,S,D)$-space) will be confined in a kind of bottleneck around the hypothetical location of $\Gamma$. This confinement will derive in a slowing down of the dynamics of these orbits and in the emergence of long transients.  

In detail, this bifurcation takes place when the nullclines given by the parallel hypersurfaces
\[
H_1: \ p\Omega(x)=\frac{\eps}{\alpha(1-\omega)} \qquad \textrm{and} \qquad
H_2: \ p\Omega(x)=\frac{\eps}{\beta},
\]
coincide at $\mu=0$ and give rise to the curve of equilibria $\Gamma$, which only exists in such case.
To determine the role of the QNC bifurcation in organising the dynamics, we study them in a neighbourhood of $\mu=0$. First, we focus on the possible $\omega$-limit sets of the orbits in $\mathring{\mathcal{U}}$ for $\mu < 0$ and $\mu > 0$. Numerical evidence discards chaotic behaviour of the system and since Proposition~\ref{PO} states the non-existence of periodic orbits, we assume hereafter that the $\omega$-limit scenarios must be either an equilibrium point of satellite extinction $Q_1 \in \{ S=0\}$ or the origin $Q_0$ itself (total extinction).
The numerical integrations to obtain the solutions of the ODEs have been performed using a Runge-Kutta-Fehlberg-Simó method\footnote{with an improvement by Prof. Lluís Alsedà} of order 7-8 with automatic step size control and local relative tolerance $10^{-15}$.

The two leading parameters conducting this study are the replication rate of the HV ($\alpha$) and that of the satRNA ($\beta$) since they establish the progeny production pace and thereby determine the persistence of the infection. On the other hand, DIPs ($D$ in the model) are a byproduct of the HV dynamics and the RdRp ($p$) plays a role subjugated to the HV. Therefore, for this reason and whenever numerical simulations are performed, $\alpha$ and $\beta$ will vary whilst the rest of parameters are assumed to take the following biologically meaningful values:
\begin{equation}\label{numerical_values}
    \kappa = 0.3, \quad \omega = 0.2, \quad \gamma = 0.6, \quad \varepsilon = 0.1, \quad \varepsilon_p = 0.01.
\end{equation}
That is, we suppose lower degradation rates for the viral types and low degradation of the RdRp in comparison, respectively, to their transcritpion and translation rates.

The term $p\,\Omega(x)$ indeed plays a central role in the dynamics, acting as an effective logistic-like function, encompassing available resources for both RNA species and the amount of the RdRp enzyme. Below, we state its value on the feasible equilibrium points.
\begin{lemma}\label{pOm_Q0}
The function $p\,\Omega(x)$ vanishes at the origin, \emph{i.e.}, ${\ds p\,\Omega(x) \big|_{Q_0} = 0}$.
\end{lemma} 

\begin{lemma} \label{lem:pOmega:Q1}
For any satRNA-extinction equilibrium point $Q_1$, we have
\[
p\,\Omega(x)\big|_{Q_1}= \frac{\eps}{\alpha(1-\omega)}.
\]
\end{lemma}

\subsubsection{Dynamics for $\mu < 0$}
Let us consider the situation previous to the {\QNC}-bifurcation (considering increasing values of $\alpha > 0$), that is, for $\mu=\alpha - \alpha^*<0$ small. Lemmas~\eqref{pOm_Q0} and~\eqref{lem:pOmega:Q1} provide insights on the possible $\omega$-limit sets in this regime by studying the term $p\,\Omega(x)$. Recall that the system does not admit periodic solutions and numerics do not show other situations like chaotic or quasiperiodic behaviour. Precisely, we have:
\begin{prop}\label{omega_limit_mu_neg}
    Let $\varphi(t,y_0)$ be the solution of~\eqref{eq1}-\eqref{eq2} with initial condition $\varphi(0,y_0) = y_0 \in \mathring{\mathcal{U}}$. Then, its $\omega$-limit $\omega(\varphi)$ satisfies that
    \begin{equation}
        \omega(\varphi) = \lim_{t\rightarrow +\infty} \varphi(t,y_0) = Q_0,
    \end{equation}
    where $Q_0=(0,0,0,0)$.
\end{prop}
\begin{remark}
This statement  also holds for orbits with initial conditions on $\{V+S+D = 1\}$, $D \in \{0,1\}$ and $p \in \{0,1\}$.
\end{remark}   
The studied system exhibits then monostability provided $y_0 \notin \{ S=0 \}$.
From $\mu<0$ it easily follows that
\[
\frac{\eps}{\beta} < \frac{\eps}{\alpha(1-\omega)}.
\]
Taking this fact into account, Proposition~\ref{omega_limit_mu_neg}, and Lemma~\ref{pOm_Q0}, we get that, for any orbit $\varphi$ with initial condition $y_0 \in\mathring{\mathcal{U}}$, there exists a minimal time value $0 \leq \tau_1 < +\infty$ such that
\begin{equation}\label{pOm<epsbeta}
    p\,\Omega(x) \Big|_{x=\varphi(t,y_0)} < \frac{\varepsilon}{\beta}, 
    \qquad \forall t > \tau_1,
\end{equation}
This  $\tau_1$ depends on the initial condition $y_0$.  Define now, for $\varphi(t,y_0)=(x(t),p(t))$, the following set:
\begin{equation}
    D_{\varphi,y_0} = \left\{0 \leq t < \tau_1 \quad \big | \quad p(t)\Omega(x(t)) = \frac{\varepsilon}{\beta} \quad \text{or} \quad p(t)\Omega(x(t)) = \frac{\varepsilon}{\alpha(1-\omega)}\right\}.
\end{equation}
This set contains all the time instants at which the orbit $\varphi(t,y_0)$ intersects either the nullcline $H_1$ or $H_2$. In the case that  $D_{\varphi,y_0}$ is non-empty we define $\tau_0 > 0$ as
\begin{equation}
\tau_0 = \sup (D_{\varphi,y_0}).
\end{equation}
Essentially, $[\tau_0,\tau_1]$ is the last time-interval in which $\varphi(t,y_0)$ is contained among both nullclines.
These definitions provide insights on the behaviour of $S(t)$ before reaching its $\omega$-limit. Precisely,
\begin{lemma}\label{S:local:maximum}
Let us assume that for an orbit $\varphi(t,y_0)=(V(t),S(t),D(t),p(t))$ with $y_0 \in \mathring{\mathcal{U}}$, the values $\tau_0$ and $\tau_1 > 0$ exist. Then $S(t)$ has a local maximum at $t=\tau_1$ and it is monotonous decreasing for all $t > \tau_1$.
\end{lemma}
Notice that, from the equation satisfied by $\dot{S}$,  $S(t)$ grows in all the time-intervals in which $p\,\Omega(x)$ falls between $\frac{\eps}{\beta}$ and $\frac{\eps}{\alpha(1-\omega)}$, and not only in the interval $(\tau_0,\tau_1)$.
Moreover, for those orbits $\varphi(t,y_0)$ with initial condition $y_0 = (V_0,S_0,D_0,p_0)$ for which we can define $\tau_0$ (\emph{i.e.}, those with non-empty $D_{\varphi,y_0}$), a lower bound for the length of the interval $(\tau_0,\tau_1)$ can be proved.

\begin{prop}[$p\,\Omega(x)$ scaling-law for $\mu<0$]
    \label{sl:mu<0:pOm}
    Let $\mu < 0$ and $\varphi$ defined as above with initial conditions satisfying $S(0),V(0)\neq 0$ and $D_{\varphi,y_0}$ non-empty. Then, there exists a constant $\tilde{K}$, which depends on the parameters of the system but independent of $\mu$, such that
    \begin{equation}
        \tau_1 - \tau_0 > \frac{\tilde{K}\alpha}{\varepsilon} \frac{1}{|\mu|}.
    \end{equation}
\end{prop}

\begin{figure}
    \centering
    \captionsetup{width=\linewidth}
    \includegraphics[width=\textwidth]{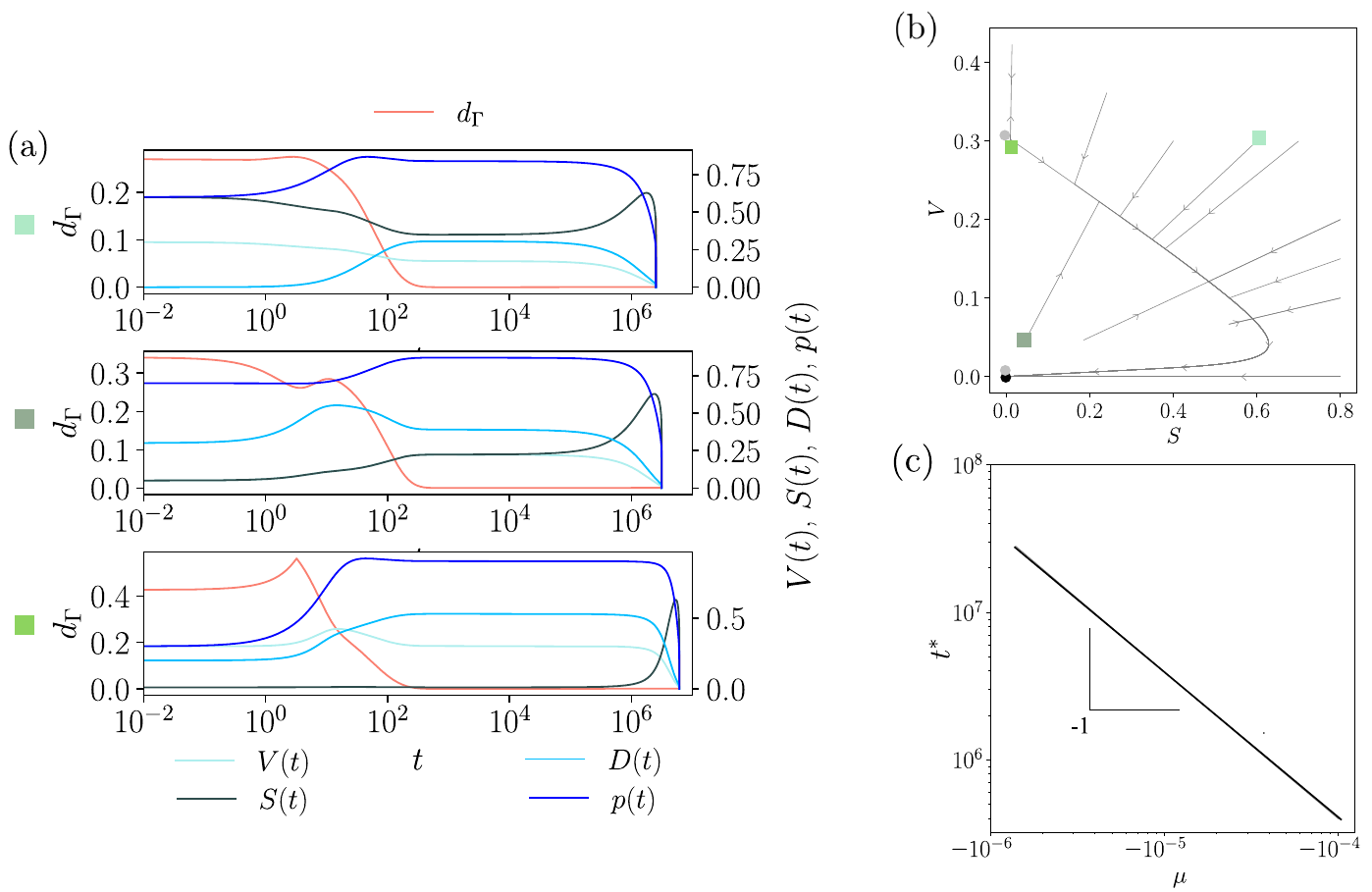}
    \caption{\small (a) Long transients around the remnant of the curve $\Gamma$ close to the {\QNC} bifurcation with $\mu = -10^{-5}$. Here the origin is a global attractor and the system becomes extinct for any initial condition with $S_0 \neq 0$. The red line shows how the distance from the orbit to the coordinates of $\Gamma$ changes along time. (b) Phase portrait with several orbits shown projected into the plane $(S,V)$. Here, the black and grey dots are local attractors and saddle points, respectively. (c) Times needed to reach a neighbourhood of the origin numerically computed as a function of the distance to the bifurcation value $\-\mu$, which follows the scaling law  $t^* \sim |\mu|^{-1}$. Here, we use $(V_0,S_0,D_0,p_0) = (0.5,0.2,0,0)$.}
    \label{alpha<alpha*}
\end{figure}

\begin{remark}\label{rmk:scaling}
In other words, $p\,\Omega(x)$ remains inside $\left( \frac{\eps}{\beta} , \frac{\eps}{\alpha(1-\omega)} \right)$ for all $t \in (\tau_0, \tau_1)$, where
\begin{equation}\label{scalinglw}
       \tau_1 - \tau_0  \sim O\left(\frac{1}{|\mu|}\right),
   \end{equation}
provided that $\tilde{K}$ is, at least, $O(\eps)$. 
%is expected to be a long time for small values of $\mu$.
\end{remark}
\begin{figure}
    \centering
    \captionsetup{width=\linewidth}
    \includegraphics[width=0.8\textwidth]{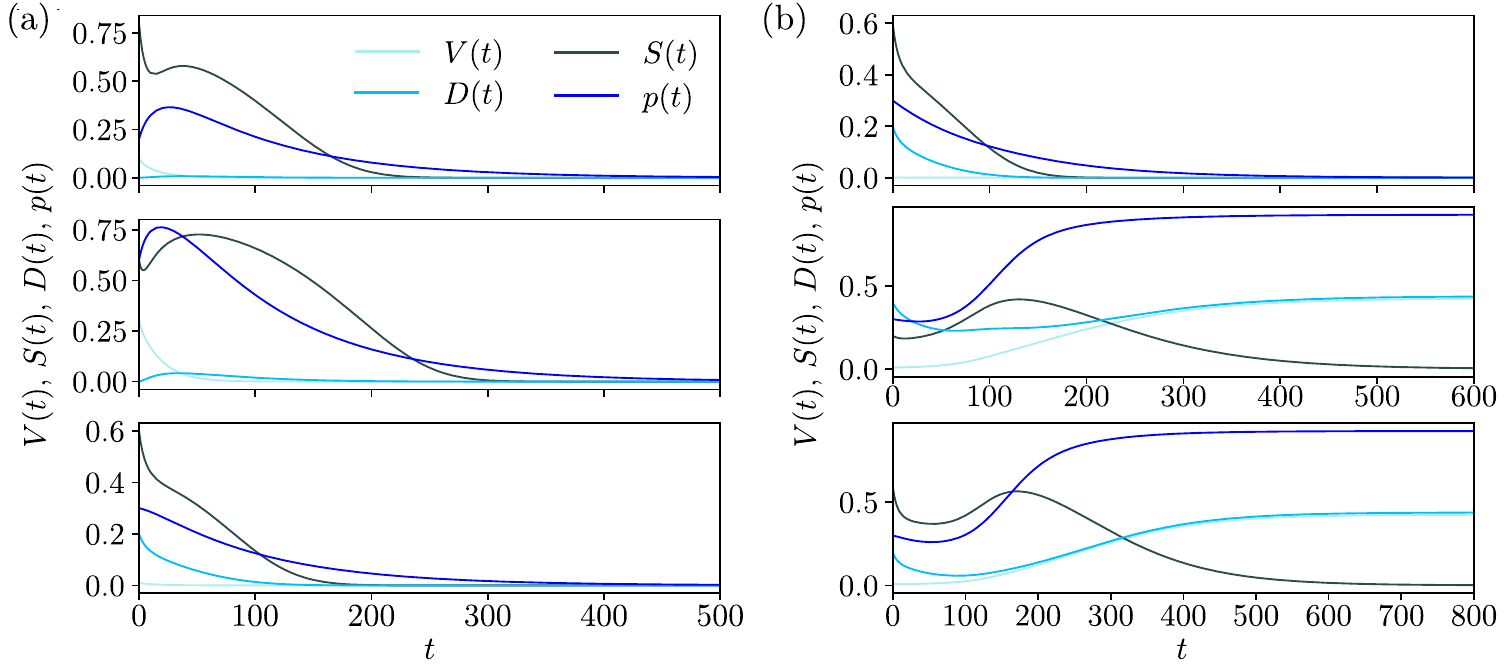}
    \caption{\small Time series for values of $\alpha$ such that $|\mu| > 0.1$ (far from the global bifurcation value), with  $\alpha < \alpha^*$ ($\mu<0$) (a); and $\alpha > \alpha^*$ ($\mu>0$) (b). Panels in (b) display the bistability scenario where both the origin and the satRNA-extinction equilibrium point $Q_1$ are locally asymptotically stable. Compare the transient times far from the bifurcation shown in this figure with those transients found close to the bifurcation (Figs.~\ref{alpha<alpha*},~\ref{pOm_TS},~\ref{mu-5_alpha>alpha*}). %Each panel corresponds to a particular set of initial conditions $y_0 = (V_0,S_0,D_0,p_0)$.
    }
    \label{ts_far}
\end{figure}
%\begin{remark}\label{rmk2}
Since the coordinate hypersurfaces $\{S=0\}$ and $\{V=0\}$ are invariant by the flow, the solution $\varphi(t,y_0)$ can never reach them  whenever its initial condition falls in 
$\mathring{\mathcal{U}}$.
%For initial conditions with $V \neq 0$, a value of $V_1 = 0$ will not be reached in finite time since $V = 0$ is an invariant manifold of the system and therefore, it can only approached in infinite time. Analogously, for initial conditions with $S \neq 0$, $S=0$ can only be reached in infinite time since $S=0$ is an invariant manifold of system (\ref{eq1})-(\ref{eq2}). For this reason, in these circumstances, we will never have $S_0 = 0$.
%\end{remark}
The scaling law for $p\Omega(x)$ stated in Proposition~\ref{sl:mu<0:pOm} can be numerically observed for all four variables, $V(t), S(t), D(t)$, and $p(t)$. Indeed, for small values of $\mu$, 
long transients appear when orbits $\varphi$ evolve close to where $\Gamma$ will be located in the case $\mu = 0$, when the {\QNC} bifurcation occurs. They behave as retained by a kind of bottleneck or ghost curve during long time intervals.  The distance of $\varphi$ to this ghost curve is represented by the red curve in Fig.~\ref{alpha<alpha*}(a) and their dynamics are illustrated in Fig.~\ref{alpha<alpha*}. Long transients in these temporal series must be compared with those in Fig.~\ref{ts_far}(a), where the time spent to reach a neighbourhood of the equilibrium $Q_0$ is five orders of magnitude smaller.

\begin{figure}
    \centering
    \captionsetup{width=\linewidth}
    \includegraphics[width=0.93\textwidth]{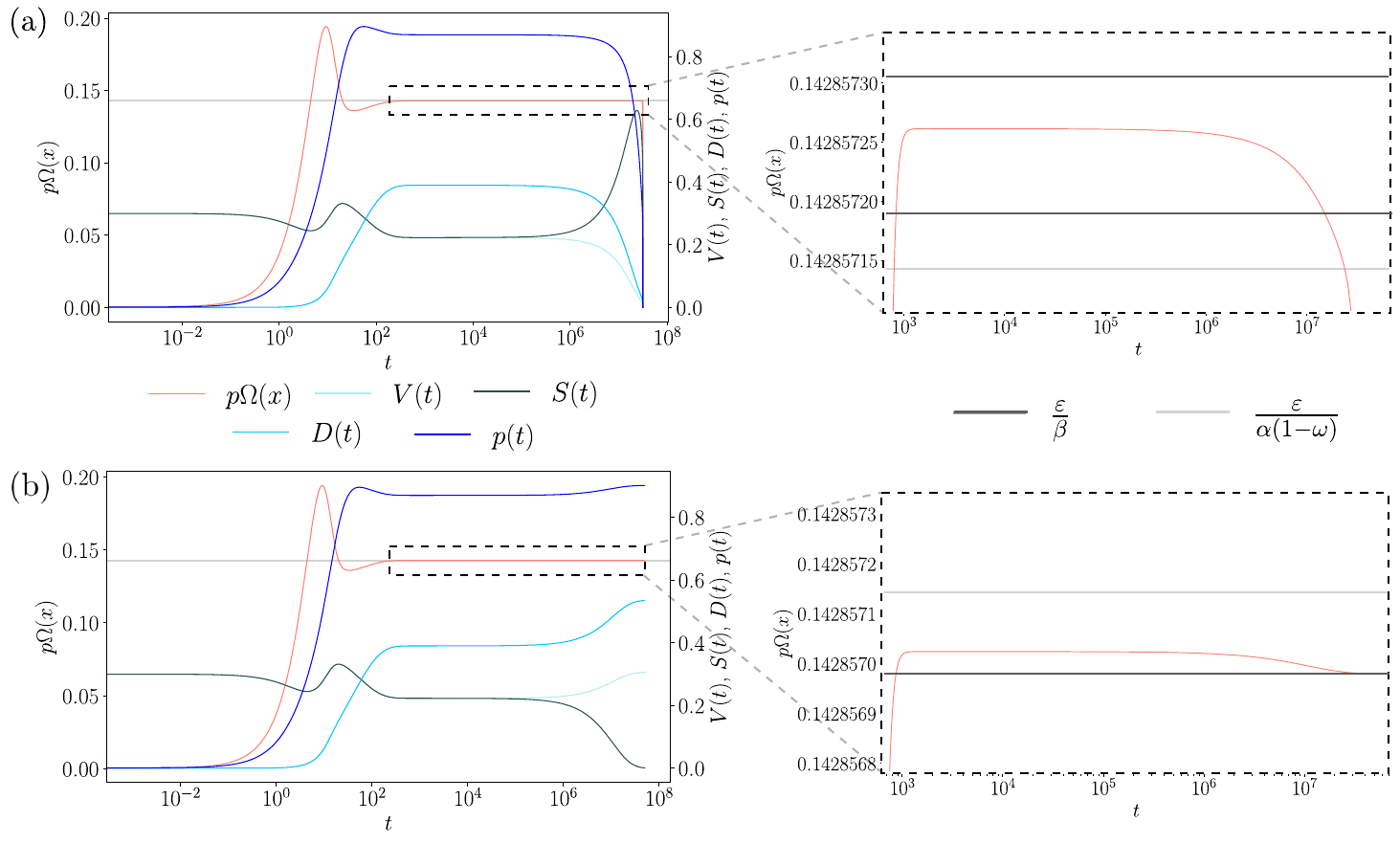}
    \caption{\small Time evolution of the four variables for $\mu \approx 0$ along with time evolution of $p\,\Omega(x)$ (red), for $\mu \lesssim 0$ (a) and $\mu \gtrsim 0$ (b). Long transients are observed for both the state variables and for $p\,\Omega(x)$ function.}
    \label{pOm_TS}
\end{figure}

Figure~\ref{alpha<alpha*}(a) also shows different transients experiencing these delays. Three different orbits with different initial conditions (green colours) are represented, also displayed in the $(V,S)$ space. Their dynamics here is characterized by a fast approach towards the bottleneck followed by a very slow passage throughout it. Numerical results have revealed that for values of $|\mu|\to 0$, the time needed to reach a fixed neighbourhood of $Q_0$ behaves like $t \sim |\mu|^{-1}$.
This scaling law is definitely determined by the proximity between the two nullclines $H_1$ and $H_2$ when $\mu<0$ is small.
Additionally, some examples of the monostable scenario provided in Proposition~\ref{omega_limit_mu_neg} are shown in Fig.~\ref{fig_processes}(b) and Fig.~\ref{alpha<alpha*}(b). Figure \ref{pOm_TS}(a) displays the time series evolution for a given orbit $\varphi$ and its corresponding value  $p\,\Omega(x)$.
%showing long transients which are consistent with the scaling in Remark \ref{rmk:scaling}.

This time scaling provides analytical and numerical results that could be related to biological chronic infections in terms of (long) transient states. Long transients involve non-zero populations for large periods of time in such a way that coexistence is allowed although not being the final stable state. However, we must notice that these dynamics may be found close to bifurcation thresholds. That is, with parameter values far from the bifurcation value $\mu=0$ the transients appear to be fast.
%with an exponential behaviour.
In the long term, since the intrinsic replication rate of the satRNA, $\beta$, is larger than the one of the HV, $\alpha$, the latter is unable to survive due to possibly a satRNA population consuming resources at a higher pace and so driving the infection to the complete extinction. The second panel in Fig.~\ref{basins_attraction}(b) reflects this behaviour in the basin of attraction of the origin. \\

The outcome differs drastically for initial conditions satisfying $S_0=0$. The invariance of this hyperplane allows equilibria state combining non all-vanishing HV, DIPs and RdRp. In this regime, when the condition for the existence of  equilibrium $Q_1$ [see Eq.~\eqref{eq:Q1:necessary:cond}] is fulfilled, there exist two non-trivial equilibrium points of this type, say $Q_1^1$ and $Q_1^2$. Their location in $\{ S=0 \}$ depends on the value of the parameter $\alpha$ and are coincident for a particular value $\alpha_*$, denoted by a grey dot in Fig.~\ref{Q1:evolution}. For $\alpha<\alpha^*$ (i.e. $\mu < 0$) both points are of saddle type: one with three attracting directions and one repelling, which is the attractor for orbits with initial conditions in $S=0$; and the other, the other way round.
The evolution of $Q_1^1$ and $Q_1^2$ for increasing values of $\alpha<\alpha^*$
is shown in Fig.~\ref{Q1:evolution}. The direction of the blue arrows correspond to growth in the value of $\alpha$.
According to the second panel in Fig.~\ref{basins_attraction}(b), only large values of $\alpha$ satisfying $\mu < 0$ allow coexistence of $V$, $D$ and $p$ inside the invariant set $\{S=0\}$.
\begin{figure}
    \centering
    \captionsetup{width=\linewidth}
    \includegraphics[width=\textwidth]{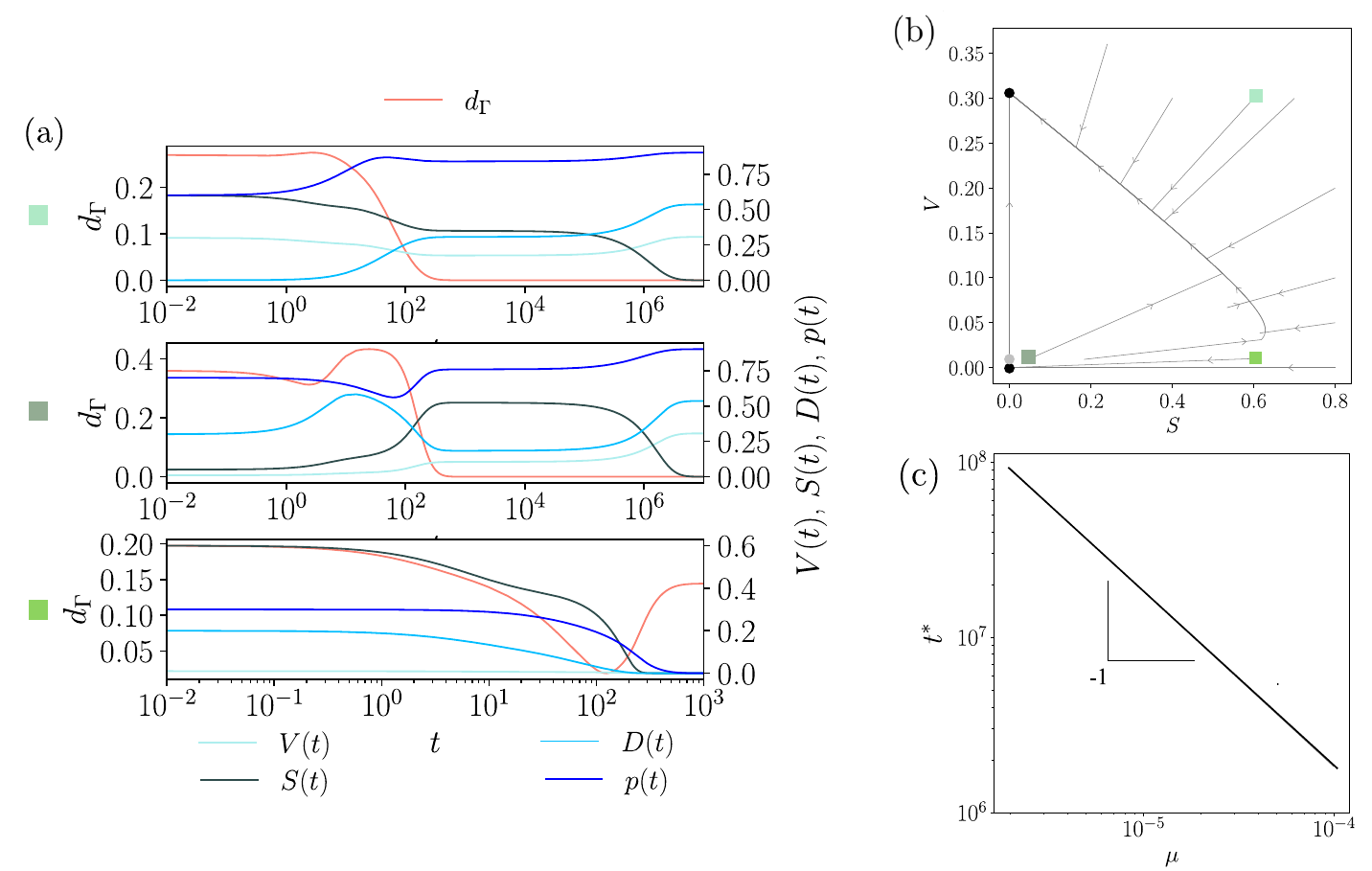}
    \caption{\small Analogous to Fig.~\ref{alpha<alpha*} with $\mu = 10^{-5}$, where the system is bistable. (a) The two upper plots show the satRNA extinction and the distance (red line) of the orbit to $\Gamma$ coordinates  (recall $\Gamma$ only exists for $\mu = 0$), for two initial conditions identified with coloured squares in the phase portrait in panel (b). The lower plot in (a) shows the full extinction since the orbit starts inside the basin of attraction of the origin. (c) The same scaling law for the transients close to bifurcation threshold is identified here numerically.}% (b) Transients and distance of the orbit (red line) to the coordinates of $\Gamma$ computed for $\mu = 0$. First and second plots display long transients when the distance to $\Gamma$ is close to zero while the third one displays fast dynamics towards the origin. (c) Necessary time to reach a neighbourhood of the origin as a function of the distance to the bifurcation value in absolute value $\-\mu$, which follows an approximate scaling law given by $t^* \sim |\mu|^{-1}$. Initial conditions used are $(V_0,S_0,D_0,p_0) = (0.5,0.2,0,0)$.}
    \label{mu-5_alpha>alpha*}
\end{figure}

%%%%%%%%%%%%%%%%%%%%%%%%%%%%%%%%%%%%%%%%%%%%%%%%%%%%%%%%%%%%%%%

\subsubsection{Dynamics for $\mu > 0$} 
In this case, the effective replication rate of $V$, given by $\alpha (1 -\omega)$, satisfies $\alpha(1-\omega) > \beta$ and so HV replicates faster than the satRNAs. Consequently, one of the two equilibria of type $Q_1$ becomes locally stable, as well as the origin. The system exhibits bistability: total extinction or a satRNA-extinction steady state
[see the phase portraits of Figs.~\ref{fig_processes}(d) and \ref{mu-5_alpha>alpha*}(b) and in the time series of Fig.~\ref{mu-5_alpha>alpha*}(a)].

For those orbits whose $\omega$-limit is an equilibrium point of type $Q_1$, the following result holds:
\begin{prop}[$Q_1$ as $\omega$-limit]
\label{prop:Q1:omega-limit}
    Consider a solution $\varphi(t,y_0)$ of system \eqref{eq1}-\eqref{eq2}, with initial conditions $y_0 = (V_0,S_0,D_0,p_0)$ and $V_0 \neq 0$, whose $\omega$-limit set is an equilibrium point of type $Q_1$. Then, for any arbitrarily small $\xi>0$ there exists a time $T=T(\xi)$ such that the associated function $p(t)\,\Omega(x(t))$ is confined in 
    \begin{equation}
        \left( \frac{\eps}{\alpha(1-\omega)} - \xi, \frac{\eps}{\alpha(1-\omega)} + \xi \right)
    \end{equation} 
    for $t > T$.
\end{prop}
So, in that case, the orbit $\varphi(t,y_0)=(x(t),p(t))$ gets trapped for infinite time in a neighbourhood of the nullcline $H_1$. In other words, it gets confined inside a bottleneck around the location where the quasi-neutral curve $\Gamma$ emerged for $\mu=0$.

\begin{figure}
    \centering
        \captionsetup{width=\linewidth} 
    \includegraphics[width=0.5\linewidth]{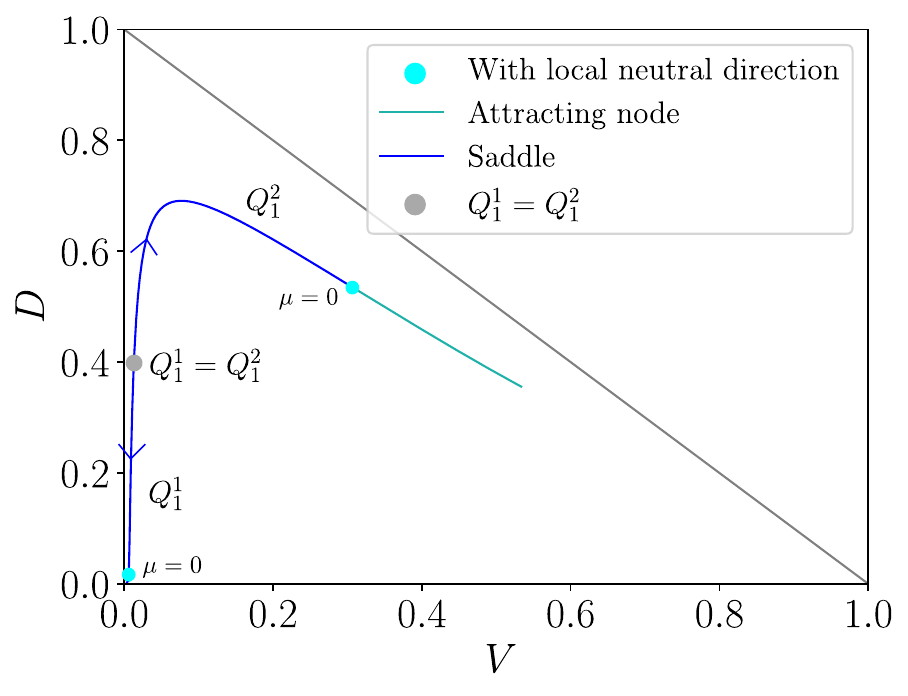}
    \caption{Evolution of the equilibrium points of type $Q_1$ and their local stability in the phase plane (V, D) at increasing $\alpha$ (and so of $\alpha(1-\omega)$) for parameters values in~\eqref{numerical_values}. The blue arrows indicate the direction of increasing values of $\alpha$. Light blue dots correspond to the case $\mu = 0$, when the system undergoes the global bifurcation. The gray line denotes the limit of the meaningful domain in the $(V,D)$ space.}
    \label{Q1:evolution}
\end{figure}

Further can be discussed about this bistability scenario with a numerical example. Consider the values of the parameters in~\eqref{numerical_values} and $\mu = 10^{-5}$. As already known,
three equilibrium points exist belonging to $\{ S=0\}$: the origin $Q_0$, which is an attracting node; $Q_1^1$, a saddle; and $Q_1^2$, also an attracting node. 
More precisely, the points $Q_1^1$ and $Q_1^2$ come from a couple of points $Q_1$ in the case $\mu=0$. 
This couple is depicted as light blue dots in Fig.~\ref{Q1:evolution} and, as end-points of the curve $\Gamma$, they have a local $1$-dim neutral direction, transversal to $\{ S=0\}$. Once the quasi-neutral curve $\Gamma$ is broken, \emph{i.e.} $\mu$ becomes positive and small, the stability of these points changes: $Q_1^2$ becomes an attracting node while $Q_1^1$ a saddle point 
(see Fig.~\ref{Q1:evolution}) with a $1$-dimensional unstable invariant manifold $W^{u}(Q_1^1)$ and a $3$-dimensional stable invariant manifold $W^{s}(Q_1^1)$. The latter is a hypersurface embedded in the $4$-dimensional space $(V,S,D,p)$ and, therefore, it splits the whole space into two disjoint subspaces, acting as a \emph{separatrix} for the basins of attraction of both points $Q_0$ and $Q_1^2$. This splitting is also preserved by projection on the $(V,S,D)$-space.
While dynamics of orbits $\varphi(t,y_0)$  have been observed numerically to be fast for those orbits with $\omega$-limit the origin, $Q_0$, [third panel in Fig.~\ref{mu-5_alpha>alpha*}(a)], for those orbits with initial conditions $y_0$ in the basin of attraction of $Q_1^2$, their dynamics appear to be slow for small $\mu>0$  and long transients are observed, as displayed in the first and second panels of Fig.~\ref{mu-5_alpha>alpha*}(a). 

\begin{figure}
    \centering
    \captionsetup{width=\linewidth}
    \includegraphics[width=\textwidth]{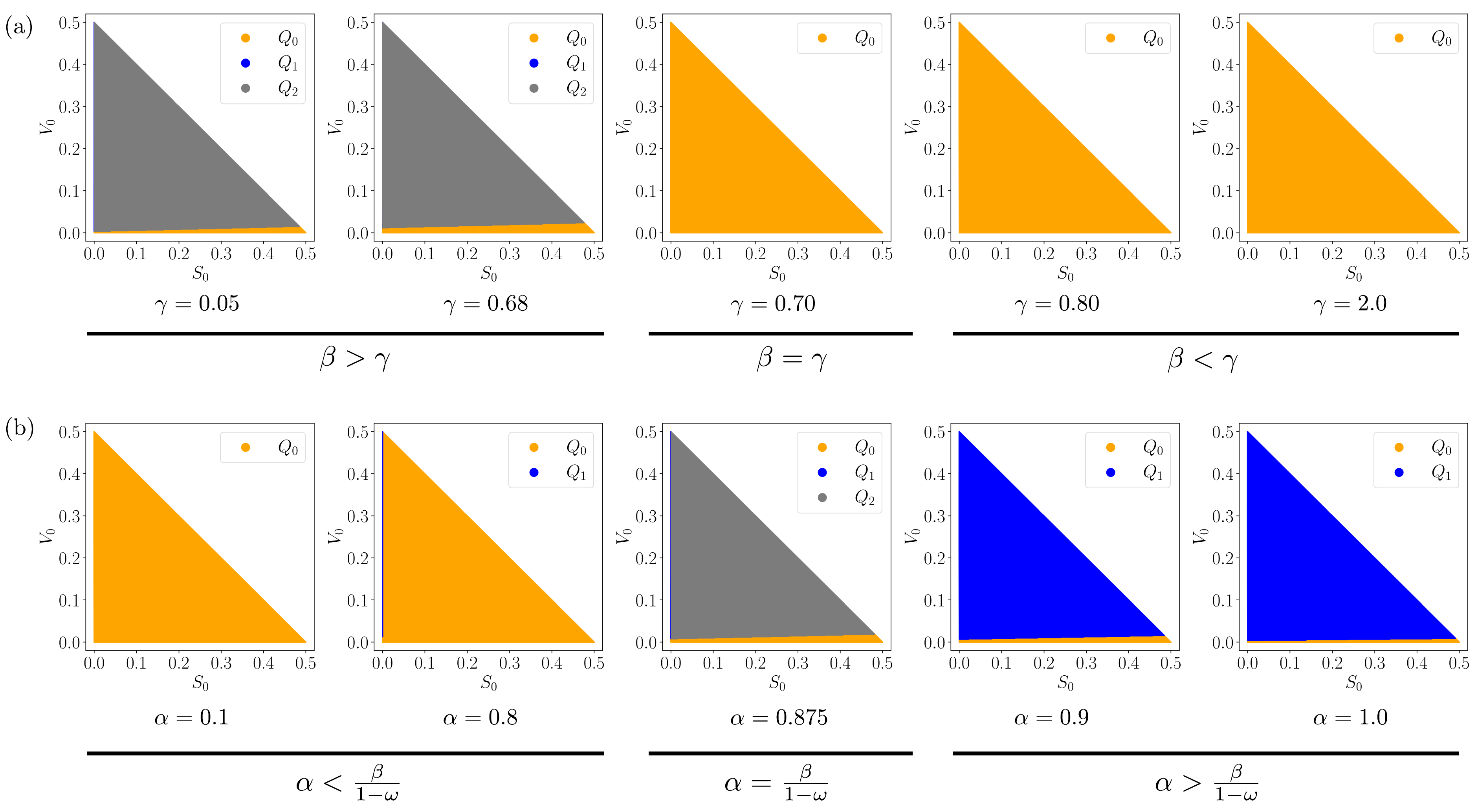}
    \caption{\small Basins of attraction for different values of the parameters in the plane of initial conditions $S_0-V_0$ (assuming HV-satRNA coinfections). The achieved $\omega$-limits for the initial conditions are shown with different colours: full extinction (orange); HV-DI-RdRp coexistence and satRNA extinction (blue); and full coexistence (grey). (a) Basins of attraction tuning $\gamma$ with $\alpha = 0.875$, $\omega = 0.2$, $\varepsilon = 0.1$, $\varepsilon_p = 0.01$, $\kappa = 0.3$ and $\beta = 0.7$. The value of $\gamma$ varies in three different regions: $\beta > \gamma$, for which the quasi-neutral curve exists; $\beta = \gamma$, critical value for which the quasi-neutral curve disappears; and $\beta < \gamma$, where only the origin and the satRNA extinction equilibrium exist. (b) Basins of attraction for different values of $\alpha$ with for $\omega = 0.2$, $\varepsilon = 0.1$, $\varepsilon_p = 0.01$, $\kappa = 0.3$, $\beta = 0.7$ and $\gamma = 0.6$.}
    \label{basins_attraction}
\end{figure}

In this scenario, a hypothetical infection with an initial low viral load of the HV and a large proportion of satRNAs may be driven to complete extinction, establishing an initial population threshold of the HV in order to achieve a persistent infection. Orbits with the origin as $\omega$-limit [see bottom panel in Fig.~\ref{mu-5_alpha>alpha*}(a) and fourth panel in Fig.~\ref{basins_attraction}(b)] are faster in reaching a vicinity of the equilibrium state since they are not affected by the \emph{ghost} of the quasi-neutral curve $\Gamma$. 
%\textcolor{red}{Thus, the border amongst basins of attraction is also the border between regions affected by the time delay.} With a view to the distance of the orbit to the coordinates of $\Gamma$, which we recall that only exists in case $\mu = 0$, we see that long transients are associated to the proximity of the orbit to the quasi-neutral curve in such a way that a remnant of $\Gamma$ causes the delay in the way towards the final stable state.
Orbits exhibiting long transients seem to satisfy also an scaling power-law for their time length of type $t \sim O(|\mu|^{-1})$. This law has been found numerically for orbits within the basin of attraction of $Q_1^2$, see Fig.~\ref{mu-5_alpha>alpha*}(c). Furthermore, the 
result in Proposition~\ref{prop:Q1:omega-limit} has been checked numerically in this case. See Fig.~\ref{pOm_TS}(b), where the term $p\,\Omega(x)$ gets trapped between the nullclines $H_1$ and $H_2$ despite the populations keep changing over time.

Finally, a numerical study of the basins of attraction of the different equilibrium points has been carried out in terms of $\gamma$ [see Fig.~\ref{basins_attraction}(a)]. A replication rate of the DIPs ($\gamma$) larger than that of the satRNA ($\beta$) leads to the total extinction of populations due to DIPs consuming resources at a higher rate. Since they are dependent on the RdRp produced by the HV, the exhaustion of resources causes the HV extinction and, consequently, it halts the production of RdRp driving the remaining RNA molecular species to extinction.

\section{Discussion}\label{discussion}
Understanding the interaction among different viral agents replicating within the same host cell is crucial to determine the fate of viral infections. A very well-studied case is the replication of a wild-type (wt) virus (also called helper virus, HV) together with its defective viral genomes (DVGs)~\cite{vignuzzi}. DVGs are deletion mutants lacking essential viral genes and are spontaneously synthesized during the regular replication of the wt virus. Some of them are usually referred as defective interfering particles (DIPs) when they interfere in the replication and virions' assembly of the HV. DIPs can reduce the symptoms caused by the HV~\cite{Dimmock2014,GALON199558,Roux1991}, modulate virulence~\cite{Cave1985} or even worsen such symptoms~\cite{turnip,Rizzeto1988}.  Considerable research has been performed on wt virus-DIPs in the last decades, including experimental~\cite{Thompson2010,Timm2014,Jaworski2017,Gribble2021,Hillung2024,Rangel2023,Munoz2025},  theoretical~\cite{Szathmary1992,Szathmary1993,Kirkwood1994,Frank2000,Sardanyes2010}, and both approaches together~\cite{Zwart2013,Chatuverdi2021,Munoz2025}.

Together with the coexistence of HV-DIPs inside the host cell, other subviral agents can coinfect or superinfect the host cells introducing further complexity and nonlinearties into the dynamics. For instance, viral satellites or satellite RNAs (satRNAs)~\cite{Palukaitis2016}. These subviral agents, which are non-related genetically to the wt virus, also kidnap the products of the HV as DIPs do, \emph{i.e.}, the RdRp and the structural proteins, competing for the cellular resources. Satellite infections, which are very common in plant viruses~\cite{Palukaitis2016,GALON199558}, can also occur in other organisms~\cite{Schmitt2000,Ribiere2010,Krupovic2016}, including humans. For example, the hepatitis delta virus requires the presence of hepatitis B virus for its spread. When coinfection of these two viruses occurs, the symptoms are much more severe accelerating hepatic cirrhosis~\cite{Rizzeto1988}.

While the dynamics of HV-DIPs have been widely investigated over the last decades~\cite{Szathmary1992,Szathmary1993,Frank2000,Sardanyes2010,Thompson2010,Rangel2023}, the impact of virus satellites or satRNAs remains less explored. Few works have investigated the dynamics for these multiple-virus systems. For instance ref.~\cite{LAZARO2024107987} investigated a mass action model for a HV synthesizing DIPs and coinfecting with a satRNA. In this manuscript we investigate a similar system including the RdRp, which introduces further nonlinearities into the system. We have identified three possible regimes as a function of parameters, mostly depending on the balance between the effective replication rate of the HV $\alpha(1-\omega)$ and the satRNA's replication rate $\beta$. Here, $\alpha$ and $\omega$ are the HV replication rate and the rate of DIPs' production, respectively. For $\alpha(1-\omega) < \beta$, the satRNA replicates faster than HV conducting the infection to its demise. For $\alpha (1-\omega) = \beta$, there exists a bistable scenario with full coexistence governed by a quasi-neutral curve of equilibria. Finally, for $\alpha(1-\omega) > \beta$, bistability is maintained driving the system towards either the satRNA  extinction or the full extinction depending on the initial conditions. 

We show that the quasi-neutral curve, which is structurally unstable, is involved in a global bifurcation that we have named as 
\textit{quasi-neutral nullcline confluence} ({\QNC}) bifurcation. At the bifurcation value, two nullcline hypersurfaces coincide giving rise to the curve of equilibria. In the neighbourhood of this bifurcation, long transients have been observed and their duration $t$ is shown to follow scaling laws of the form $t \sim |\mu|^{-1}$, with $\mu = \alpha - \alpha^*$, $\alpha^*$ being the bifurcation value associated to the critical replication rate of the HV. Such scaling law has been found numerically and derived analytically. This scaling exponent coincides with the one obtained in a host-parasite model with a quasi-neutral curve of equilibria~\cite{Fontich2022}. Close to the bifurcation, the orbits are strongly conditioned by a ghost quasi-neutral curve, and an apparent state of persistent HV-DIPs-satRNAs infection arises.

As mentioned, ref.~\cite{LAZARO2024107987} investigated the dynamics of a similar model without including the RdRp. This model revealed three different scenarios depending on parameter values: ($i$) full extinction; ($ii$) HV-DIPs coexistence with satRNA clearance, and ($iii$) HV-DIPs-satRNA coexistence. This system revealed a very narrow region of bistability allowing for scenarios ($ii$) and ($iii$) to coexist. Our results indicate that the explicit inclusion of the RdRp involves wider bistability scenarios, and full coexistence only governed by the quasi-neutral curve. Dynamics governed by quasi-neutral manifolds have been identified in Lotka-Volterra competition models~\cite{Lin2012}, in strains’ competition models of disease dynamics~\cite{Kogan2014}. Within the field of theoretical virology, quasi-neutral lines have been investigated in simple models of asymmetric RNA replication~\cite{Sardanyes2018}. More recently, planes of equilibria have been identified in epidemiological-like models for coronaviruses infection in cell cultures~\cite{Munoz2025}.

Our results suggest that HV-DIPs-satRNA coexistence at early stages of replication could be governed by quasi-neutral manifolds or, more likely, by transients near the global bifurcation of the quasi-neutral curve. It is known that virus replication can be spatially structured within the cell. In this sense, viral replication factories could play a key role in promoting further coexistence of these viral agents~\cite{Romero2014}. Future research should focus on the impact of space on these quasi-neutral manifolds. To the extend of our knowledge, this topic has not been yet addressed. Also, the impact of noise on these neutral manifolds and on the identified scaling laws close to bifurcation threshold may be worth to explore within the context of coinfections and superinfections caused by viral satellites.

\section*{Acknowledgements}
O.L has been supported by the predoctoral program AGAUR-FI ajuts (2023 FI-1 00354) Joan Oró, which is backed by the Secretariat of Universities and Research of the Department of Research and Universities of the Generalitat of Catalonia, as well as the European Social Plus Fund.
J.T.L has been supported by the project PID2021-122954NB-I00 funded by MCIU/AEI/10.13039/501100011033/ and “ERDF a way of making Europe”, and by the grant “Ayudas para la Recualificación del Sistema Universitario Español 2021-2023”. J.T.L also thanks the Laboratorio Subterráneo de Canfranc, the I2SysBio and the Institut de Mathématiques de Jussieu-Paris Rive Gauche (Sorbonne Université) for their hospitality as hosting institutions of this grant. We also thank the MCIU/AEI/10.13039/ 501100011033/, through the María de Maeztu Program for Units of Excellence in R\&D (CEX2020-001084-M) and CERCA Programme/Generalitat de Catalunya for institutional support.
J.S has been also supported by the Ramón y Cajal grant RYC-2017-22243 funded by MCIU/AEI/10.13039/ 501100011033 and “ESF invests in your future”.
S.F.E was supported by grant PID2022-136912NB-I00 funded by MCIN/AEI/ 10.13039/501100011033 and by “ERDF a way of making Europe”, and by Generalitat Valenciana grant CIPROM/2022/59.
We want to thank the Department of Mathematics and Computer Science from Universitat de Barcelona for providing us with the 7th-8th order Runge-Kutta-Fehlberg algorithm.

\appendix
\addcontentsline{toc}{section}{Appendices}
\input{appendix}

\bibliographystyle{unsrt}
%\bibliography{ReferencesCQNE}

\end{document}

%% file: appendix.tex
\section{Equilibrium points and general dynamics}
\label{se:appendix:1}
\subsection{Proof of Lemma~\ref{p_equilibrium}}

    Equation~\eqref{eq2} at the equilibrium reads as $\kappa V^*(1-p^*) - \varepsilon_p p^* = 0$. Straightaway, it leads to expression
    \begin{equation}
        p^* = p^*(V^*) = 1 - \frac{\varepsilon_p}{\kappa V^* + \varepsilon_p}.
    \end{equation}
    On the one hand, $\kappa V^* + \varepsilon_p \geq \varepsilon_p$, since $\kappa > 0$, $\varepsilon_p > 0 $ and $V^* \geq 0$. Therefore, $0 \leq p^* < 1$ and $p^*(V^*) = 0$ if and only if $V^* = 0$.

\subsection{Proof of Proposition~\ref{PO}}
We prove the result by \textit{reductio ad absurdum}. Let us assume there exists a solution of system (\ref{eq1})-(\ref{eq2}) that is $T$-periodic with period $T > 0$. This implies that all the variables of the solution must be $T$-periodic, \emph{i.e.}, $V(t)$, $S(t)$, $D(t)$ and $p(t)$ are $T$-periodic. Focusing on $p(t)$, we can rewrite equation~\eqref{eq2} as
\begin{equation}\label{non_homo_p}
    \dot{p}(t) = -(\kappa V(t) + \varepsilon_p) p(t) + \kappa V(t),
\end{equation}
which is a non-homogeneous 1st order linear differential equation that can be expressed as $\dot{p}(t) = a(t) p + b(t)$, where $a(t) = -(\kappa V(t) + \varepsilon_p)$ and $b(t) = \kappa V(t)$. General solution for this linear differential equation may be written as
\begin{equation}
    p(t) = p_0 \exp\left(\int_{0}^{t} a(s) ds\right) + \int_{0}^{t} b(r) \exp\left(\int_{r}^{t} a(s) ds \right) dr,
\end{equation}
and, plugging the values of $a(t)$ and $b(t)$, we obtain:
\begin{equation}
    p(t) = p_0 \exp\left(-\varepsilon_p t - \kappa \int_{0}^{t} V(s) ds \right) + \kappa \exp(-\varepsilon_p t) \int_{0}^{t} V(r) \exp\left(\varepsilon_p r - \kappa \int_{r}^{t} V(s) ds \right) dr.
\end{equation}
Defining
\begin{equation}\label{w(t)}
    w(t) = p_0 \exp\left(-\kappa \int_{0}^{t} V(s) ds\right) + \kappa \int_{0}^{t} V(r) \exp\left(\varepsilon_p r - \kappa \int_{r}^{t} V(s) ds \right) dr,
\end{equation}
we can rewrite $p(t)$ as
\begin{equation}
    p(t) = \exp\left(-\varepsilon_p t\right) w(t).
\end{equation}
Since $p(t)$ is $T$-periodic, we have $p(0) = p(T)$ where $p(0) = p_0$ and $p(T) = \exp(-\varepsilon_p T) w(T)$. Then,
\begin{equation*}
    p_0 = e^{-\varepsilon_p T} w(T) \Longleftrightarrow w(T) = p_0 e^{\varepsilon_p T}.
\end{equation*}
Introducing the expression of $w(t)$ given by equation \eqref{w(t)} in the last equality and rearranging terms, we obtain:
\begin{equation}
    p_0 \left(e^{\varepsilon_p T} - \exp\left(-\kappa \int_{0}^{T} V(s) ds\right)\right) = \kappa \int_{0}^{T} V(r) \exp\left(\varepsilon_p r - \kappa \int_{r}^{T} V(s) ds \right) dr.
\end{equation}
We now consider two cases:
\begin{itemize}
    \item Consider
    \begin{equation}
        e^{\varepsilon_p T} - \exp\left(-\kappa \int_{0}^{T} V(s) ds\right) \neq 0.
    \end{equation}
    Then $p_0$ has the following expression:
    \begin{equation}
        p_0 = \frac{\kappa}{e^{\varepsilon_p T} - \exp\left(-\kappa \int_{0}^{T} V(s) ds\right)} \int_{0}^{T} V(r) \exp\left(\varepsilon_p r - \kappa \int_{r}^{T} V(s) ds \right) dr.
    \end{equation}
    This equation uniquely determines the initial value $p(0) = p_0$ that makes $p(t)$ to be $T$-periodic but there exist infinitely many possible values for $p_0$ on the periodic solution. Therefore, we have a contradiction. 
    \item Consider now 
    \begin{equation}
        e^{\varepsilon_p T} - \exp\left(-\kappa \int_{0}^{T} V(s) ds\right) = 0.
    \end{equation}
    Therefore,
    \begin{equation}
        \frac{1}{T} \int_{0}^{T} V(s) ds = -\frac{\varepsilon_p}{\kappa},
    \end{equation}
    which provides an expression for the mean value of function $V(t)$ in the interval $[0,T]$. However, by hypothesis we are considering initial conditions in $\mathcal{U}$ and, since this domain is positively invariant, $V(t)$ is a positive function within it. Therefore, its mean value can not be negative.
\end{itemize}
Both cases lead to contradictions and, therefore, $p(t)$ can not be periodic. Since there is a component in solution $(V(t), S(t), D(t), p(t))$ which is not $T$-periodic, solution can not be $T$-periodic.

\subsection{Proof of Lemma~\ref{eq:V:zero}}
For $V^*=0$ we have $\dot{V} = 0$. Assuming it, equation $\dot{p} = 0$ only has $p^* = 0$ as a valid solution. Conditions $V^* = 0$ and $p^* = 0$ lead equation $\dot{D} = 0$ only to be satisfied for $D^* = 0$. Finally, $p^*= 0$ leads equation $\dot{S} = 0$ to be satisfied only when $S^* = 0$. Thus, for $V^* = 0$, the only equilibrium point possible is $Q^* = Q_0 = (0,0,0,0)$. 

\subsection{Proof of Lemma~\ref{gamma:alpha:Q0}}
Equilibrium equation for $V$ with the constraint $\alpha(1-\omega) = \gamma$ reads:
    \begin{equation} 
        V^* \left(\gamma p \Omega(x) - \varepsilon\right) = 0, 
    \end{equation}
    which gives rise to two different cases: $V^*=0$ or $p\,\Omega(x) = \varepsilon/\gamma$. Case $V^*=0$ is already described in Lemma~\ref{eq:V:zero}. Assume then $p\,\Omega(x) = \varepsilon/\gamma$. Equilibrium equation for $D$ reads:
    \begin{equation}
        (\omega \alpha V^* + \gamma D^*) \frac{\varepsilon}{\gamma} - \varepsilon D^* = 0 \Longleftrightarrow \frac{\omega \alpha \varepsilon}{\gamma} V^* = 0.
    \end{equation}
    Then, necessarily, one must have $V^*=0$. Lemma~\ref{eq:V:zero} completes the proof. 

\subsection{Proof of Lemma~\ref{Q0:local:stability}}
The differential matrix of system \eqref{eq1}-\eqref{eq2} evaluated at the origin is given by 
    \begin{equation}
        DF({\bf{0}}) = \left(
        \begin{matrix}
            -\varepsilon & 0 & 0 & 0 \\
            \kappa & -\varepsilon_p & 0 & 0 \\
            0 & 0 & -\varepsilon & 0 \\
            0 & 0 & 0 & -\varepsilon
        \end{matrix}
        \right),
    \end{equation}
    with eigenvalues $\lambda_1 = -\varepsilon_p$ and $\lambda_2 = \lambda_3 = \lambda_4 = -\varepsilon$. Since we are assuming $\varepsilon_p,\varepsilon > 0$, all eigenvalues are real and negative and, therefore, the origin is locally asymptotically stable for all values of parameters in the Lyapunov sense.

\subsection{Proof of Proposition~\ref{Q1_prop}}
    Assuming $V\neq 0$ and $S = 0$, equilibrium for $V$ is satisfied if
    \begin{equation}\label{cond:V:Q1}
        V \left(p\Omega(x) - \frac{\varepsilon}{\alpha(1-\omega)} \right) = 0 \Longleftrightarrow p\Omega(x) = \frac{\varepsilon}{\alpha (1-\omega)},
    \end{equation}
    which automatically introduces restriction $\alpha > \varepsilon/(1-\omega)$ since $p \leq 1$ and $\Omega(x) \leq 1$ and we only have $p\Omega(x) = 1$ for $p=1$ and $V=D=S=0$, which is biologically impossible. From the equilibrium equation for $D$, and taking into account equilibrium condition for $V$, one gets
    \begin{equation}\label{cond:D:Q1}
        D(V) = \frac{\alpha\omega}{\alpha(1-\omega) - \gamma} V > 0 \Longleftrightarrow \alpha (1-\omega) > \gamma.
    \end{equation}
    Thus, in order to obtain biologically meaningful solution values, by virtue of the former considerations, we must have
    \begin{equation}\label{par:cond:Q1}
        \alpha (1-\omega) > \mathrm{max}\{\varepsilon,\gamma\}.
    \end{equation}
    From (\ref{cond:V:Q1}), using expressions for $D$ and $p$ as functions of $V$ given, respectively, by (\ref{cond:D:Q1}) and (\ref{pequilib}) and after some trivial algebraic manipulations, one gets
    \begin{equation}
        \varepsilon\kappa \frac{\alpha - \gamma}{\alpha (1-\omega) - \gamma} V^2 - \varepsilon \kappa \left(1 - \frac{\varepsilon}{\alpha (1-\omega)}\right) V + \frac{\varepsilon^2\varepsilon_p}{\alpha (1-\omega)} = 0,
    \end{equation}
    which gives rise to a 2nd-degree polynomial in variable $V$ and real coefficients $q_1(V) = T_2 V^2 + T_1 V + T_0$. Coefficients are labelled as follows: 
    \begin{equation}
        T_2 = \frac{\varepsilon \kappa (\alpha - \gamma)}{\alpha (1-\omega) - \gamma}, \qquad T_1 = - \kappa \varepsilon \left(1 - \frac{\varepsilon}{\alpha(1-\omega)}\right), \qquad T_0 = \frac{\varepsilon_p \varepsilon^2}{\alpha (1-\omega)}.
    \end{equation}
    By the Fundamental Theorem of Algebra, $q_1(V)$ has exactly two roots in $\mathbb{C}$. From condition (\ref{par:cond:Q1}), we get $\alpha > \gamma$ since $1-\omega < 1$ and thus, $T_2 > 0$. We also have $T_0 > 0$. Again, condition (\ref{par:cond:Q1}) leads to $T_1 < 0$. By the Descartes' rule of signs, since there are exactly two changes of sign in the coefficients of $q_1(V)$, it contains either 2 or 0 real and positive roots. Defining the discriminant of $q_1(V)$ as $\Delta = T_1^2 - 4T_2 T_0$, we distinguish cases below:
    \begin{enumerate}
        \item[a)] If $\Delta > 0$, since $q_1(-V)$ is a second degree polynomial with real and positive coefficients, the number of negative roots of $q_1(V)$ is zero by virtue of the Descartes' rule of signs. Then, $\Delta > 0$ necessarily leads to the existence of two real and positive roots for $q_1(V)$.
        \item[b)] If $\Delta = 0$, then
        \begin{equation}
            V = \frac{-T_1}{2T_2} > 0,
        \end{equation}
        which is a real and positive root with multiplicity 2.
    \end{enumerate}
    Then, cases a) and b) give, at most, two real and positive roots $V_1$ of $q_1(V)$. For each of these roots, there exist values of $p$ and $D$ given, respectively, by (\ref{pequilib}) and (\ref{cond:D:Q1}).

\subsection{Proof of Proposition~\ref{Q2_prop}}
Assuming $S_2 \neq 0$, the equilibrium equation $\dot{S} = 0$ reads
    \begin{equation}
        S_2 \left(\beta p_2 \Omega(x_2) - \varepsilon\right) = 0,
    \end{equation}
    where $p_2$ is given by (\ref{pequilib}) according to the equilibrium equation for $p$ and $x_2 = (V_2,S_2,D_2)$. This equation leads necessarily to condition $p_2 \Omega(x_2) = \varepsilon / \beta$. By substituting this expression in the equilibrium equation $\dot{V} = 0$,
    \begin{equation*}
        V_2 \left(\alpha (1-\omega) \frac{\varepsilon}{\beta} - \varepsilon \right) = 0.
    \end{equation*}
    Since $V_2 \neq 0$, this equation is only fulfilled for $\alpha (1 -\omega) = \beta$. In particular, for $\alpha (1-\omega) \neq \beta$ there are no coexistence equilibrium points. 
    Assuming both conditions, the equilibrium equation $\dot{D} = 0$ reads
    \begin{equation}
        \omega \alpha V_2 = (\beta - \gamma) D_2 \Longrightarrow D_2 (V_2) = \frac{\omega \alpha}{\beta - \gamma} V_2 = \frac{\beta \omega}{(\beta-\gamma)(1-\omega)} \, V_2,
    \end{equation}   
    since $V_2 \neq 0$ and $\alpha (1-\omega)= \beta$. The expression above is only biologically meaningful if $\beta > \gamma$ and using the expressions $p_2(V_2)$ and $D_2(V_2)$, it follows that
    \begin{eqnarray}
        p_2 \Omega(x_2) =\frac{\varepsilon}{\beta} 
        &\Leftrightarrow& \left(1 - \frac{\varepsilon_p}{\kappa V_2 + \varepsilon_p}\right) \left(1 - V_2 - \frac{\omega \alpha}{\beta - \gamma} V_2 - S_2\right) = \frac{\varepsilon}{\beta} \nonumber \\
        &\Leftrightarrow& -\kappa \left(1 + \frac{\omega \alpha}{\beta - \gamma}\right) V_2^2 + \kappa \left(1 - S_2 - \frac{\varepsilon}{\beta}\right) V_2 - \frac{\varepsilon\varepsilon_p}{\beta} = q_2(V_2) = 0. \label{conic:eq}
    \end{eqnarray}
    Therefore, in order for $V_2 \in (0,1]$ to be biologically meaningful, it must be a root of the polynomial $q_2(V_2)$ defined above. The discriminant $\Delta_2$ of $q_2(V_2)$ is given by:
    \begin{equation*}
        \Delta_2 = \kappa^2 \left(1 - S_2 - \frac{\varepsilon}{\beta}\right)^2 - 4\kappa \left(1 + \frac{\alpha \omega}{\beta - \gamma}\right) \frac{\varepsilon \varepsilon_p}{\beta}.
    \end{equation*}
    Condition $\Delta_2 \geq 0$ provides real solutions for the roots of polynomial $q_2(V_2)$. Then, by solving this inequality in terms of $S_2$, we have $\Delta_2 \geq 0$ if and only if 
    \begin{equation*}
        S_2 \in I = \left(-\infty, 1 - \frac{\varepsilon}{\beta} - 2 \sqrt{\frac{\varepsilon \varepsilon_p}{\kappa \beta} \left(1 + \frac{\alpha \omega}{\beta - \gamma}\right)}\right) \cup \left(1 - \frac{\varepsilon}{\beta} + 2 \sqrt{\frac{\varepsilon \varepsilon_p}{\kappa \beta} \left(1 + \frac{\alpha \omega}{\beta - \gamma}\right)}, + \infty \right).
    \end{equation*}
    The value of $S_2$ obtained will be biologically meaningful if and only if $S_2 \in I \cap (0,1)$.
    Then, for each value of $S_2 \in I \cap (0,1)$, the corresponding values of $V_2(S_2)$ are given by the roots of $q_2(V_2)$ in the interval $(0,1)$. For each of the biologically meaningful values of $V_2(S_2)$, the corresponding values of $D_2(S_2)$ and $p_2(S_2)$ are obtained according to their respective expressions. Then, coexistence equilibrium points are given by the 1-parametric family
    \begin{equation*}
        \mathcal{E} \coloneqq \left\{(V_2,S_2,D_2,p_2) = (V_2(S_2), S_2, D_2(S_2), p_2(S_2)) \in \mathcal{U} \quad \big| \ S_2 \in I \cap (0,1), \ q_2(V_2)=0 \right\},
    \end{equation*}
    which corresponds to those values of $(V_2,S_2)$ belonging to the conic~\eqref{conic:eq} that fall into $\mathring{\mathcal{U}}$.

\subsection{Proof of Proposition~\ref{D:invariant:surface}}
We define the following function:
    \begin{equation}
        H \left(V(t),D(t)\right) = D(t) - \frac{\beta\omega}{(\beta - \gamma)(1-\omega)} V(t).
    \end{equation}
    Then the surface we want to prove to be invariant is given by $H(V(t),D(t)) = 0$. Using equations \eqref{eq1} and \eqref{eq5}, we have:
    \begin{align*}
        \frac{dH(V(t),D(t))}{dt} &= \frac{\partial H}{\partial V} \frac{dV}{dt} + \frac{\partial H}{\partial D} \frac{dD}{dt} = -\frac{\beta\omega}{(\beta-\gamma)(1-\omega)} \dot{V} + \dot{D} \\
        &= (\alpha\omega V + \gamma D) p\,\Omega(x) - \varepsilon D - \frac{\beta\omega}{(\beta - \gamma)(1-\omega)} \left(\alpha(1-\omega) V p\,\Omega(x) - \varepsilon V\right).
    \end{align*}
    Evaluating the expression above in the surface $H(V(t),D(t)) = 0$, we obtain:
    \begin{align*}
        \frac{dH(V(t),D(t))}{dt} &= \left(\alpha \omega V + \gamma \frac{\beta\omega}{(\beta - \gamma)(1-\omega)} V\right) p\,\Omega(x) - \varepsilon \frac{\beta \omega}{(\beta - \gamma)(1 - \omega)} V \\
        &\hspace{0.5cm}- \frac{\beta\omega}{(\beta - \gamma)(1-\omega)} \alpha(1-\omega) V p\,\Omega(x) + \varepsilon \frac{\beta\omega}{(\beta - \gamma)(1-\omega)} V \\
        &= p\,\Omega(x) V \left(\alpha \omega + \frac{\beta\omega\gamma}{(\beta-\gamma)(1-\omega)} - \frac{\beta\omega \alpha (1-\omega)}{(\beta - \gamma)(1-\omega)}\right) \\
        &= \frac{p\,\Omega(x) V}{(\beta - \gamma)(1-\omega)} \left(\beta \omega \alpha (1-\omega) - \gamma \omega \alpha (1-\omega) + \beta \omega \gamma - \beta \omega \alpha (1-\omega)\right) \\
        &=0,
    \end{align*}
    where the last equality follows from hypothesis $\alpha (1-\omega) = \beta$. Therefore, $H(V(t),D(t)) = 0$ is invariant for \eqref{eq1}-\eqref{eq2}.

\subsection{Proof of Proposition~\ref{first:integral:prop}}
Assuming $\alpha(1-\omega) = \beta$, equations \eqref{eq1} and \eqref{eq4} read, respectively, as:
    \begin{equation}
        \dot{V}=\alpha(1-\omega) V \left( p\,\Omega(x) - \frac{\eps}{\alpha(1-\omega)} \right) \quad \text{and} \quad \dot{S}=\alpha(1-\omega) S \left( p\,\Omega(x) - \frac{\eps}{\alpha(1-\omega)} \right),
    \end{equation}
    and therefore, assuming $V \neq 0$, one has
    \begin{equation*}
        \dot{S}V - S\dot{V} = 0 \Longleftrightarrow \frac{\dot{S}V - S\dot{V}}{V^2} = 0 \Longleftrightarrow \frac{d}{dt} \left(\frac{S}{V}\right) = 0,
    \end{equation*}
    which provides the condition for $S(t)/V(t)$ to be a first integral.

\section{The global bifurcation}
\label{se:appendix:2}

\subsection{Proof of Lemma~\ref{lem:pOmega:Q1}}
    From Proposition~\ref{Q1_prop} we know that
    \[
        p_1=p_1(V_1)=1-\frac{\eps_p}{\kappa V_1 + \eps_p}.
    \]
    Since equilibrium points $Q_1$ satisfy $p\,\Omega(x) = \varepsilon/(\alpha(1-\omega))$, we can express the value of $D_1$ as
    \[
        D_1(V_1) = 1 - V_1 - \frac{\varepsilon (\kappa V_1 + \varepsilon_p)}{\kappa \alpha V_1 (1-\omega)}.
    \]
    Substituting these values into $p\Omega(x)$ we get
    \[
    p\Omega(x)\big|_{Q_1} = \left( 1-\frac{\eps_p}{\kappa V_1 + \eps_p} \right) \left( \frac{\eps(\kappa V_1 + \eps_p)}{\alpha \kappa V_1(1-\omega)} \right) = \frac{\eps}{\alpha(1-\omega)}.
    \]

\subsection{Proof of Proposition~\ref{omega_limit_mu_neg}}
    Since system~\eqref{eq1}-\eqref{eq2} is analytic, $\mathcal{U}$ is compact and positively invariant by the flow and for $\mu \neq 0$ there are no periodic orbits,  equilibrium points exists in the interior $\mathring{\mathcal{U}}$ and numerical evidence discards the existence of invariant torus or quasi-periodic orbits; the only possibility for the $\omega$-limit set of $\varphi(t,y_0)$ is to be equilibrium point on $\{S = 0\}$. By \textit{reductio ad absurdum}, assume $\omega(\varphi) = Q_1$. From Lemma~\eqref{lem:pOmega:Q1} we have:
    \begin{equation}
        \lim_{t \rightarrow +\infty} p\Omega(x)\Big|_{(V,S,D,p) = \varphi(t,y_0)} = \frac{\varepsilon}{\alpha (1-\omega)} > \frac{\varepsilon}{\beta},
    \end{equation}
    where the last inequality follows from condition $\mu < 0$. Therefore, there exists $+\infty > t_1 > 0$ such that
    \begin{equation}
        p\Omega(x) \Big|_{(V,S,D,p) = \varphi(t,y_0)} > \frac{\varepsilon}{\beta}, \quad \forall t > t_1.
    \end{equation}
    Consequently, variable $S(t)$ in $\varphi(t,y_0)$ satisfies
    \begin{equation}
        \dot{S(t)} = \beta S(t) \left(p\Omega(x) - \frac{\varepsilon}{\beta}\right) > 0, \quad \forall t > t_1,
    \end{equation}
    that is to say, $S(t)$ is strictly increasing within interval $(t_1,+\infty)$, which contradicts the fact that $\omega(\varphi) = Q_1$. Notice that $S(t_1) \neq 0$ since $S=0$ is an invariant manifold and we can not reach it in finite time for initial conditions in $\mathring{\mathcal{U}}$. Thus, necessarily $\omega(\varphi) = Q_0$.

\subsection{Proof of Lemma~\ref{S:local:maximum}}
    Assume $\mu < 0$ and, consequently, $\alpha < \alpha^*$. Then, for $\tau_0 < t < \tau_1$, we have:
    \begin{align*}
        \frac{\varepsilon}{\beta} < p \Omega(x) < \frac{\varepsilon}{\alpha (1-\omega)} &\Longleftrightarrow 0 < p\Omega(x)- \frac{\varepsilon}{\beta} < \frac{\varepsilon}{\alpha(1-\omega)} - \frac{\varepsilon}{\beta} = \frac{\varepsilon}{\alpha\beta} (\alpha^* - \alpha) \\
        &\Longleftrightarrow 0 < \beta \left(p\Omega(x) - \frac{\varepsilon}{\beta} \right) < \frac{\varepsilon}{\alpha} (\alpha^* - \alpha) \\
        &\Longleftrightarrow 0 < \frac{\dot{S}}{S} < \frac{\varepsilon}{\alpha} (\alpha^* - \alpha).
    \end{align*}
    Since $S > 0$ in $\mathcal{U}$, we need $\dot{S} > 0$ in order for the last inequality to be fulfilled. Therefore, for $\tau_0 < t < \tau_1$, $S(t)$ is monotonous increasing and for $t > \tau_1$, according to definition of $\tau_1$, $p\Omega(x) < \varepsilon/\beta$. In this case,
    \begin{align*}
        p\Omega(x) < \frac{\varepsilon}{\beta} \Longleftrightarrow \beta\left(p\Omega(x) - \frac{\varepsilon}{\beta}\right) < 0 \Longleftrightarrow \frac{\dot{S}}{S} < 0.
    \end{align*}
    By the same argument as above, we need $\dot{S} < 0$ and thus $S(t)$ is monotonous decreasing for $t > \tau_1$. Then necessarily, $S(t)$ must have a local maximum at $t = \tau_1$.

\subsection{Proof of Proposition~\ref{sl:mu<0:pOm}}
    Since $\mu < 0$, we have $\alpha < \alpha^*$. Then, for all $\tau_0 < t < \tau_1$, we have
    \begin{align*}
        \frac{\varepsilon}{\beta} < p \Omega(x) < \frac{\varepsilon}{\alpha (1-\omega)} &\Longleftrightarrow 0 < p\Omega(x)- \frac{\varepsilon}{\beta} < \frac{\varepsilon}{\alpha(1-\omega)} - \frac{\varepsilon}{\beta} = \frac{\varepsilon}{\alpha\beta} (\alpha^* - \alpha) \\
        &\Longleftrightarrow 0 < \beta \left(p\Omega(x) - \frac{\varepsilon}{\beta} \right) < \frac{\varepsilon}{\alpha} (\alpha^* - \alpha) \\
        &\Longleftrightarrow 0 < \frac{\dot{S}}{S} < \frac{\varepsilon}{\alpha} (\alpha^* - \alpha).
    \end{align*}
    Integrating the expression above in the interval $(\tau_0,\tau_1)$, we get:
    \begin{align}\label{sl:S:mu<0}
        0 < \int_{\tau_0}^{\tau_1} \frac{\dot{S}}{S} dt < \frac{\varepsilon}{\alpha} (\alpha^* - \alpha) \int_{\tau_0}^{\tau_1} dt \Longleftrightarrow 0 < \log \left(\frac{S(\tau_1)}{S(\tau_0)}\right) < \frac{\varepsilon}{\alpha} (\alpha^* - \alpha) (\tau_1 - \tau_0).
    \end{align}
    Denoting by $S_1 = S(\tau_1)$ and $S_0 = S(\tau_0)$, we have $S_1 > S_0 > 0$. Analogously, for $t \in (\tau_0,\tau_1)$, we can reconstruct the evolution equation for $V$ and integrating the resulting expression in the inequality, we get:
    \begin{equation}\label{sl:V:mu<0}
        0 < \log\left(\frac{V(\tau_0)}{V(\tau_1)}\right) < \frac{\varepsilon}{\alpha^*} (\alpha^* - \alpha) (\tau_1 - \tau_0),
    \end{equation}
    where we again denote $V_0 = V(\tau_0)$ and $V_1 = V(\tau_1)$ that satisfy $V_0 > V_1 > 0$. Then, from inequalities (\ref{sl:S:mu<0}) and (\ref{sl:V:mu<0}), we obtain, respectively:
    \begin{align*}
        \tau_1 - \tau_0 &> \frac{\alpha}{\varepsilon} \frac{1}{\alpha^* - \alpha} \log\left(\frac{S_1}{S_0}\right), \\
        \tau_1 - \tau_0 &> \frac{\alpha^*}{\varepsilon} \frac{1}{\alpha^* - \alpha} \log\left(\frac{V_0}{V_1}\right) > \frac{\alpha}{\varepsilon} \frac{1}{\alpha^* - \alpha} \log\left(\frac{V_0}{V_1}\right).
    \end{align*}
    Defining now
    \begin{equation}
        \tilde{K} \coloneqq \text{max} \left\{\log\left(\frac{V_0}{V_1}\right), \log\left(\frac{S_1}{S_0}\right)\right\},
    \end{equation}
    which does not depend on $\mu$, we get our result proved taking into account that $|\mu| = \alpha^* - \alpha$:
    \begin{equation}
        \tau_1 - \tau_0 > \frac{\tilde{K}\alpha}{\varepsilon} \frac{1}{|\mu|}.
    \end{equation}

\subsection{Proof of Proposition~\ref{prop:Q1:omega-limit}}
    Let us define 
    \begin{equation}
        \delta_1 = \frac{\varepsilon}{\beta} \quad \text{and} \quad \delta_2 = \frac{\varepsilon}{\alpha (1-\omega)}.
    \end{equation}
    Assuming $Q_1 = (V_1,0,D_1,p_1)$ as the $\omega$-limit of an orbit $\varphi(t,y_0)$, there exists $\xi > 0$ arbitrarily small and $t_0(\xi) > 0$ such that, for all $t > t_0$
    \begin{equation}\label{pOm_mu}
        \left|p\Omega(x) - \delta_2\right| < \xi,
    \end{equation}
    since $Q_1$ is found at the intersection between one of the components of $\dot{V} = 0$ nullcline ($p\Omega(x) = \delta_2$) and the invariant subset $S=0$, according to Lemma~\ref{lem:pOmega:Q1}. Expression (\ref{pOm_mu}) follows directly from the definition of $\omega$-limit. Straightforwardly, 
    \begin{equation}
        \delta_2 - \xi < p \Omega(x) < \delta_2 + \xi < \delta_1,
    \end{equation}
    and $p\Omega(x)$ gets infinitely confined.